\documentclass[3p, onecolumn]{elsarticle}
\usepackage[utf8]{inputenc}

\makeatletter
\def\ps@pprintTitle{%
 \let\@oddhead\@empty
 \let\@evenhead\@empty
 \def\@oddfoot{}%
 \let\@evenfoot\@oddfoot}
\makeatother

\usepackage{amsmath,amssymb}
\usepackage[version=4]{mhchem}  
\usepackage{siunitx}
\usepackage{physics}
\usepackage{ulem}

\usepackage{graphicx}

\usepackage{hyperref}

\usepackage[inline]{enumitem}
\usepackage{caption}
\usepackage{subcaption}

\newcommand{\kin}[1]{k \in \{\text{#1}\}}
\newcommand{\degC}{^\circ \mathrm{C}}
\newcommand{\mrs}{\mathrm{s}}
\newcommand{\mre}{\mathrm{e}}
\newcommand{\mrn}{\mathrm{n}}
\newcommand{\mrp}{\mathrm{p}}
\newcommand{\micro}{\mathrm{micro}}

\DeclareMathOperator{\arcsinh}{arcsinh}
\newcommand{\dl}{^\text{dl}}

\title{A Continuum of Physics-Based Lithium-Ion Battery Models Reviewed}

\author[1,13]{F. Brosa Planella\corref{cor}}
\author[2,3,13]{W. Ai}
\author[4,13]{A. M. Boyce}
\author[5,6,13]{A. Ghosh}
\author[7,13]{I. Korotkin}
\author[8,13]{S. Sahu}
\author[9]{V. Sulzer}
\author[10,13]{R. Timms}
\author[4,13]{T. G. Tranter}
\author[12,13]{M. Zyskin}
\author[2,13]{S. J. Cooper}
\author[5,13]{J. S. Edge}
\author[8,13]{J. M. Foster}
\author[5,13]{M. Marinescu}
\author[2,13]{B. Wu}
\author[7,13]{G. Richardson} 

\address[1]{WMG, University of Warwick, Gibbet Hill Road, Coventry, CV4 7AL, United Kingdom}
\address[2]{Dyson School of Design Engineering, Imperial College London, London, SW7 2AZ, United Kingdom}
\address[3]{School of Civil Engineering, Southeast University, 211189, China}
\address[4]{Department of Chemical Engineering, University College London, London, WC1E 7JE, United Kingdom}
\address[5]{Department of Mechanical Engineering, Imperial College London, London, SW7 2AZ, United Kingdom}
\address[6]{Department of Chemical Engineering and Technology, Indian Institute of Technology (BHU), Varanasi, Uttar Pradesh 221005, India}
\address[7]{Mathematical Sciences, University of Southampton, University Road, Southampton, SO17 1BJ, United Kingdom}
\address[8]{School of Mathematics and Physics, University of Portsmouth, Lion Terrace, Portsmouth, PO1 3HF, United Kingdom}
\address[9]{Carnegie Mellon University, 5000 Forbes Ave, Pittsburgh, PA 15213, United States}
\address[10]{Mathematical Institute, University of Oxford, Oxford, OX2 6GG, United Kingdom}
\address[12]{Department of Engineering Science, University of Oxford, Parks Road, Oxford, OX1 3PJ, United Kingdom}
\address[13]{The Faraday Institution, Quad One, Becquerel Avenue, Harwell Campus, Didcot, OX11 0RA, United Kingdom}

\cortext[cor]{Corresponding author: Ferran.Brosa-Planella@warwick.ac.uk}

\begin{document}

\begin{abstract}
Physics-based electrochemical battery models derived from porous electrode theory are a very powerful tool for understanding lithium-ion batteries, as well as for improving their design and management. Different model fidelity, and thus model complexity, is needed for different applications. For example, in battery design we can afford longer computational times and the use of powerful computers, while for real-time battery control (e.g. in electric vehicles) we need to perform very fast calculations using simple devices. For this reason, simplified models that retain most of the features at a lower computational cost are widely used. Even though in the literature we often find these simplified models posed independently, leading to inconsistencies between models, they can actually be derived from more complicated models using a unified and systematic framework. In this review, we showcase this reductive framework, starting from a high-fidelity microscale model and reducing it all the way down to the Single Particle Model (SPM), deriving in the process other common models, such as the Doyle-Fuller-Newman (DFN) model. We also provide a critical discussion on the advantages and shortcomings of each of the models, which can aid model selection for a particular application. Finally, we provide an overview of possible extensions to the models, with a special focus on thermal models. Any of these extensions could be incorporated into the microscale model and the reductive framework re-applied to lead to a new generation of simplified, multi-physics models.
\end{abstract}

\maketitle

\tableofcontents



\section{Introduction}\label{sec:introduction}

Rechargeable batteries are used in a variety of applications, spanning many scales in terms of stored energy, from portable consumer electronics (Wh), through electric vehicles (kWh) and up to grid scale energy storage (MWh). At present, lithium-ion batteries dominate the market, due to their high power and energy densities. Ongoing research seeks not only to improve battery performance and affordability, but also to extend their lifetime, while enhancing both safety and sustainability.

Modelling is one of the key tools to enable these improvements to lithium-ion batteries. A model is simply an abstract representation of an object or system, which can be used to gain understanding and make predictions. For batteries, these models usually take the form of mathematical equations, together with appropriate boundary and initial conditions, and they allow us to make quantitative predictions of the battery's behaviour. The main advantage of models over experiments is that they are cheaper and faster to run, and therefore allow the testing and validation of many more hypotheses (and prototypes) than an experiment with equivalent resources. Models can be applied to many different challenges that arise throughout the entire battery life: from the discovery of new materials to the development of battery management systems. In this review, we focus on models that describe the cycling of a battery, as it is charged and discharged. Such models have multiple applications in battery design and control.

The basic electrochemical unit of a battery is the single cell. This consists of a pair of porous electrodes (negative and positive) separated by a porous separator, all sandwiched between metallic current collectors, as shown in Figure \ref{fig:battery_sketch}. The entire porous region between the current collectors is typically flooded with a liquid electrolyte, containing dissolved lithium ions. Both positive and negative electrodes are composed of agglomerations of microscopic particles of active materials into which lithium ions can (de)intercalate. These particles are connected together via a porous network of polymer binder, typically doped with a carbon additive that improves both the mechanical integrity and electronic conductivity of the electrodes. During discharge, the higher chemical affinity of lithium ions to the positive electrode active material, compared to that of the negative electrode, causes lithium ions to de-intercalate at the surface of negative particles into the electrolyte and flow through the pores of the separator, before intercalating into one of the positive particles. The resulting transfer of charge from the negative electrode to the positive electrode leads to a potential difference between these two entities, which can be used to drive an electronic current through an external circuit.

Most people are familiar with the common lithium-ion battery formats (such as cylindrical, prismatic and pouch batteries) used in consumer electronics, and that also form the basic building blocks of the large battery packs used in high power applications, such as electric vehicles. These batteries, as illustrated in Figure \ref{fig:battery_sketch}, are made of multiple single cells stacked on top of each other (pouch), or of a large single cell wound into a compact configuration (cylindrical and prismatic), in order to provide high density energy storage and power.

\subsection{Overview of the different modelling approaches}
A lithium-ion battery is a complex device whose performance depends on a diverse set of physical and chemical phenomena, interacting over a disparate range of scales. As a consequence, a wide variety of different modelling approaches can be adopted to investigate different aspects of its behaviour \cite{Howey2020}. These extend from atomistic models, used to understand how materials behave at the nanoscale, to equivalent-circuit models, used in battery control algorithms. While quantum mechanical, atomistic models play a key role in, for example, electrode material discovery, they are limited to very small length scales (typically 100s of atoms) and short time scales, see \cite{Prentice2020}. As a consequence of these limitations, the dynamics at the cell level are usually simulated using a continuum model. These models, which are the main focus of this review, consider the different components of the battery as continuous media, rather than as discrete particles or atoms. This assumption gives rise to models that can handle larger length and time scales. Moving up to even larger length scales, such as the battery or pack level, we encounter system models which focus on describing the joint behaviour of multiple cells or batteries. These often build on continuum models, and, because pack models are typically based on multiple versions of a continuum model coupled together, they have driven the need for simple, yet accurate continuum models. Beyond these pack models, we can find techno-economic analyses, which require time scales on the order of years.

Continuum battery models generally fall into two categories: empirical and physics-based. Empirical models focus on incrementally adjusting equations and their parameters to find the best fit for experimental data, representing the observable behaviour of the battery. Equivalent-circuit models (ECMs) are a family of empirical models widely used in battery management systems (BMS), because of their computational speed and small parameter set. However, since ECMs are entirely phenomenological and not based on the underlying physics, they cannot shed light on the internal mechanisms of the battery, operate outside the regime in which they are parameterised, or be relied upon to predict long-term battery behaviour. In contrast, physics-based models represent the physical phenomena underpinning battery behaviour, and can be used both to produce highly accurate simulations of battery performance and to interrogate its internal behaviour. Hybrid models exist too, which combine the best aspects of the two approaches, so as to reach a suitable compromise in the trade-off between accuracy and speed.

\begin{figure}
    \centering
    \includegraphics[width=\linewidth]{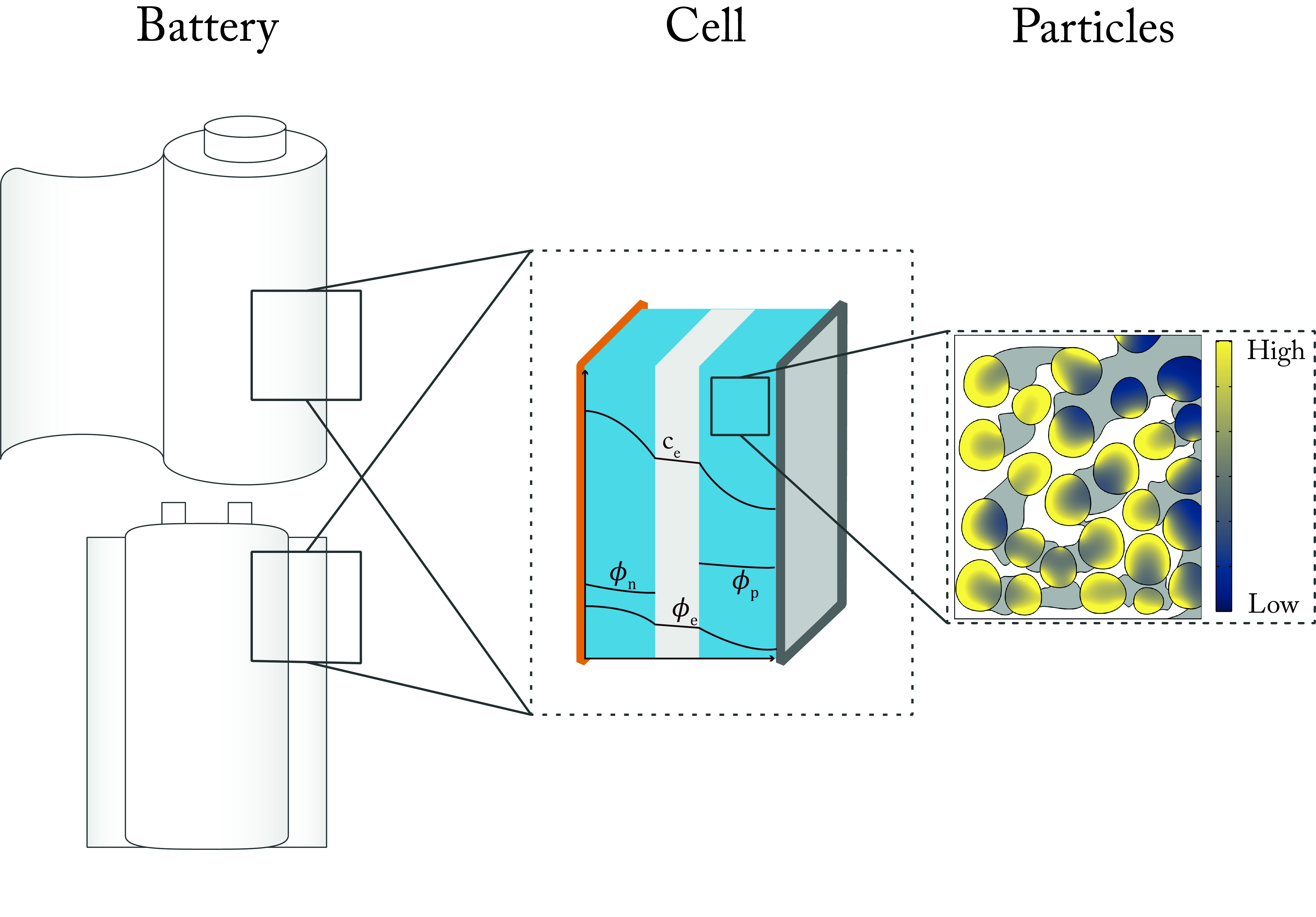}
    \caption{Sketch of the battery components at different scales. A battery is composed of multiple layers of single cells, which in turn are made of a porous structure, composed of electrode particles (yellow/black) held together by a conductive binder (grey), filled with a liquid electrolyte (white). The diagram also illustrates the electrochemical variables in the model: ion concentration in the electrolyte $c_\mre$, electrolyte potential $\phi_\mre$, electrode potentials $\phi_\mrn$ and $\phi_\mrp$, and concentration of intercalated lithium $c_\mrn$ and $c_\mrp$ (yellow/black colourmap).}
    \label{fig:battery_sketch}
\end{figure}

Physics-based continuum, electrochemical battery models were initially developed in the 1960's \cite{Newman1962} and have since been adapted to a range of battery chemistries, including lead-acid \cite{Newman1975}, nickel/metal hydride \cite{Paxton1997}, lithium-air \cite{Liu2016}, and lithium-ion \cite{Doyle1993,Fuller1994,Fuller1994a,Newman2004}. The latter is commonly referred to as the Doyle-Fuller-Newman (DFN) model and it has dominated battery continuum modelling since the early 1990’s. Reductions of the aforementioned physics-based models started with the single particle model proposed by Atlung \textit{et al.} \cite{Atlung1979} and was later extended by Prada \textit{et al.} \cite{prada2012simplified} to include the lithium-ion distribution in the electrolyte. Single Particle Models (SPMs) assume that all the particles within an electrode can be represented by a single spherical particle, and thus significantly reduce the complexity of the model. These physics-based, electrochemical models also provide insight into the behaviour of the multiple internal variables, as depicted in Figure \ref{fig:battery_sketch}, which cannot easily be measured in an \textit{in operando} set-up. In particular, they describe the potential and current distributions in both the porous electrodes and the electrolyte, the lithium ion concentration within the electrolyte and the distribution of intercalated lithium within the electrode particles. When properly implemented and calibrated, these models can provide fast and accurate predictions of real batteries, which have many possible applications. For example, they can be used as a design tool, in order to facilitate new electrode, cell and pack architectures and to assess their potential performance, thereby minimising the need for expensive prototyping. Model simulations can also be used to determine which of the many types of batteries available on the market best fit a particular use case. Other possible uses include: thermal management, \textit{in operando} control, battery monitoring and lifetime estimation.

\subsection{How to navigate this article?}

The aim of this review is not only to present a suite of physics-based battery models, but also to show how each model can be derived from a more complex one, by following a systematic approach based on some underlying assumptions. Here we do not provide the full details of each derivation but instead refer to the relevant works in the literature. Rather, our aim is to provide a framework that puts the relationships between the various physics-based models in context, and simplifies the task of finding (and reading) the relevant papers. It is hoped that this will enable readers to derive reduced models which incorporate additional physics, by re-applying the framework we describe to the appropriate microscale model. In order to suit the needs of readers from different backgrounds the review is structured so that each section can be read independently from the others. However, to get a complete picture we recommend reading the sections in the order that they are presented.

The most complex model presented in this article is the microscale model, and this serves as the starting point for our discussion. Indeed, by commencing from the microscale model, we can systematically derive all subsequent models in order of decreasing complexity: the homogenised model, the Doyle-Fuller-Newman (DFN) model, the Single Particle Model with electrolyte dynamics (SPMe) and the Single Particle Model (SPM). For convenience, we have split the reduction process into two sections: from microscale to DFN model (Section \ref{sec:full_to_DFN}) and from DFN model (via SPMe) to SPM (Section \ref{sec:DFN_to_SPM}). Given the central role that the DFN model plays in battery modelling, Section \ref{sec:full_to_DFN} is used to derive it from first principles, while Section \ref{sec:DFN_to_SPM} outlines how the DFN model can be further simplified to obtain the Single Particle Models. For each model in turn, we discuss the assumptions needed to derive them from their parent model and pose the appropriate differential equations, boundary conditions and initial conditions, using notation that is consistent across all models. After presenting each model, we discuss their relative merits and disadvantages in Section \ref{sec:comparison}. We accomplish this by considering the models in order of increasing complexity (starting from the SPM and working up to the microscale model), in order to discuss the extra insight and predictive capability that can be gained by choosing a more complex model over a simpler one, and the associated computational cost that must be paid in order to achieve this. The aim of making these comparisons is to aid the reader in the choice of a suitable model for their application of interest. In Section 5, we introduce thermal models and show how to couple them to electrochemical models. Finally, in Section \ref{sec:extensions}, we discuss different directions in which the models could be extended. Each of these extensions could fill its own review paper, so in this article, they are presented briefly with references for further reading.

\begin{figure}
    \centering
    \includegraphics[width=0.9 \linewidth]{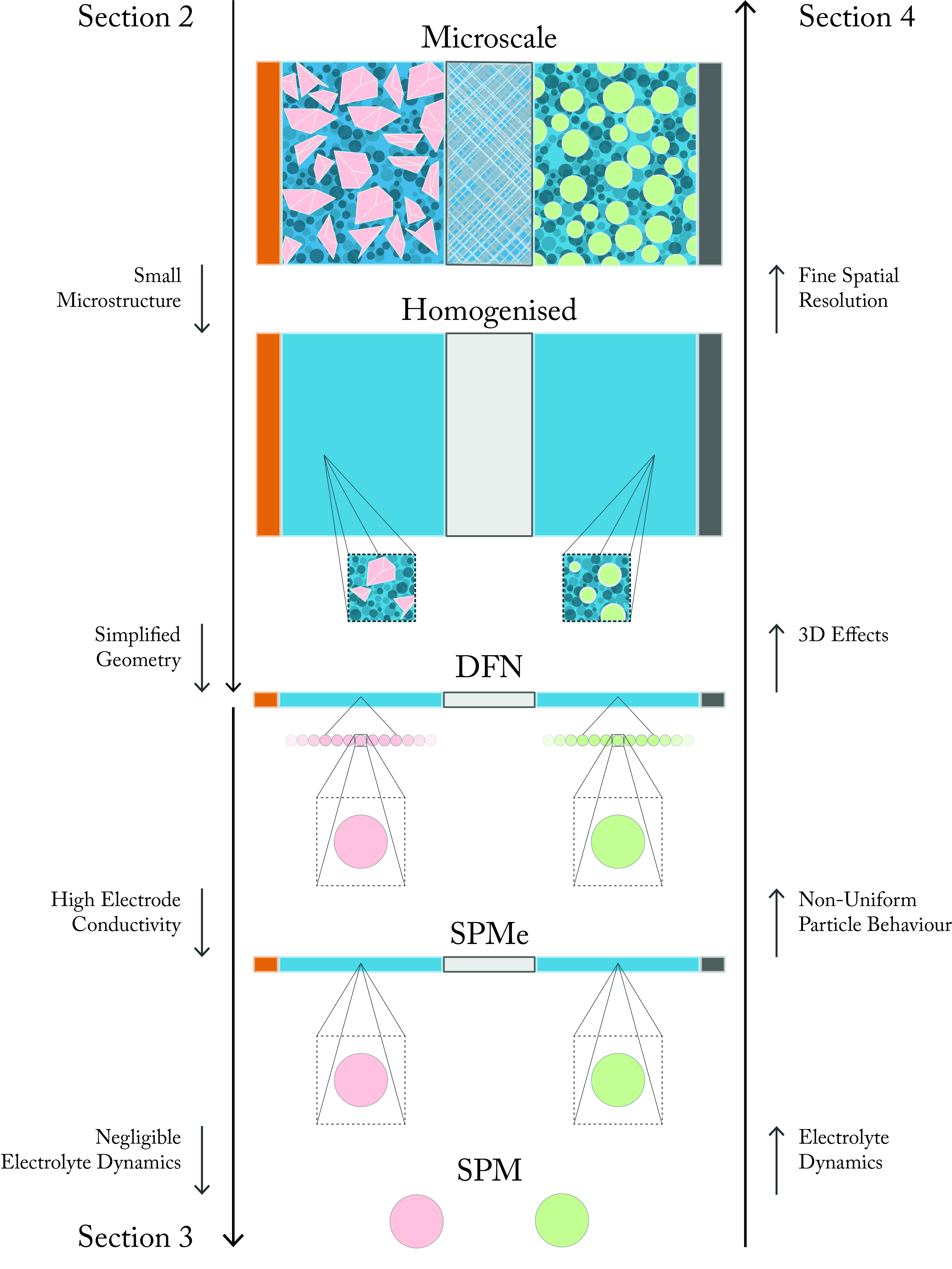}
    \caption{Graphical summary of this article. The models range from higher (top) to lower (bottom) complexity. The reduction process is separated into two sections: Section \ref{sec:full_to_DFN} encompasses the reduction from the microscale model to the DFN model, while Section \ref{sec:DFN_to_SPM} encompasses the reduction from the DFN model to the SPM. Section \ref{sec:comparison} presents a bottom-up discussion of the models.}
    \label{fig:summary}
\end{figure}

\section{Microscale model to Doyle-Fuller-Newman model}\label{sec:full_to_DFN}

The focus of Sections \ref{sec:full_to_DFN} and \ref{sec:DFN_to_SPM} is on physics-based continuum models of a single cell (as illustrated in Figure \ref{fig:sketch_full}). These models consider the processes occurring in the electrode particles and the electrolyte but do not explicitly consider effects outside this region (for example, in the current collectors). However, as we shall discuss in Section \ref{sec:extensions}, the models can be readily extended to configurations (such as pouch, cylindrical and prismatic batteries) in which these external effects play a significant role (see \cite{Timms2021}).

Continuum physics-based battery models are built around conservation laws for lithium ions and the negative counterion species in the electrolyte, and for lithium ions in the electrode particles. These conservation laws are supplemented by appropriate transport laws (also termed constitutive equations) that relate the ion fluxes to the gradients of the ion species electrochemical potentials. An equation for the electric potential in electrolyte is obtained by enforcing charge neutrality and the intercalation reaction rates on the surfaces of the electrode particles are determined from the difference between the  electrochemical potential of a lithium ion on the electrode particle surface and one in the adjacent electrolyte via the Butler-Volmer equation. It is also usual to account for the electrical resistance of the matrix of electrode binder and electrode particles, which comprise the solid structure that forms each electrode, by writing down a law for the conservation of current and coupling this to Ohm's law, in the relevant domains.

\subsection{Microscale model}
\label{sec:microscale}


\begin{figure}
    \centering
    \includegraphics[width=\linewidth]{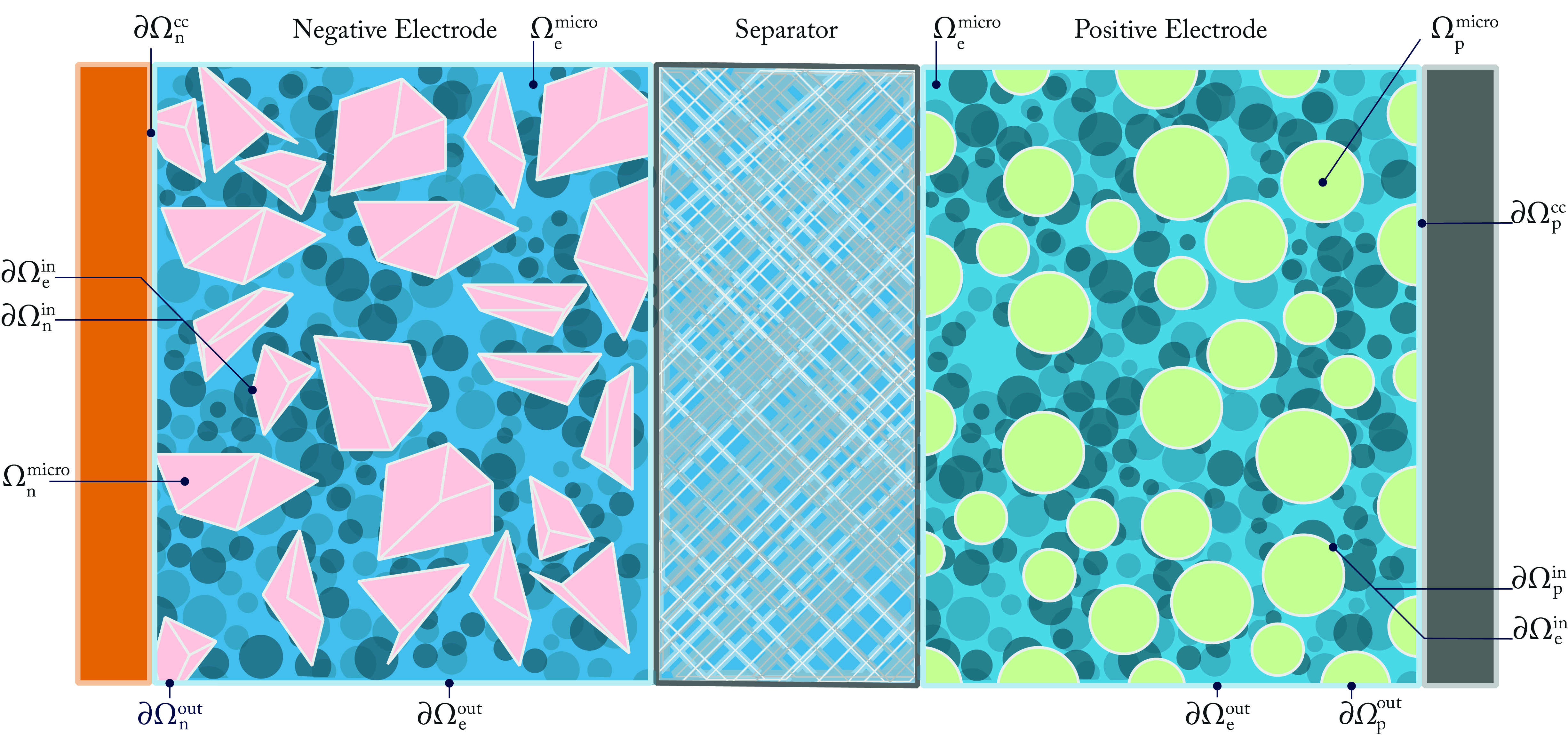}
    \caption{Sketch of the geometry of the microscale model. Each porous electrode ($\Omega_\mrn^\micro$ and $\Omega_\mrp^\micro$) is composed of a matrix that includes both active and inactive materials. The voids in this porous structure $\Omega_\mre^\micro$ are occupied by the electrolyte. The boundary of each domain is split into different subsets, so we can impose distinct boundary conditions.}
    \label{fig:sketch_full}
\end{figure}

The first model that we discuss, which we term the microscale model, provides a continuum description of charge transport down to the scale of single electrode particles. The microscale model underpins all other models discussed in this work and, indeed, as shown in Sections \ref{sec:full_to_DFN} and \ref{sec:DFN_to_SPM}, can be used to derive them. Sub-nanometer length scale models, such as those for quantum mechanics and molecular dynamics, are not explicitly considered here but can, in principle, be used to derive parameters for the physics-based models discussed here.

Since the microscale model is based on electrode geometries resolved down to the scale of individual electrode particles, an accurate representation of the microstructure is required for the model to be utilised to its full potential. A common approach is to reconstruct model microstructures from image data. This is, however, a very challenging problem due to the disparity in scales of the features that need to be resolved, which range from around 100 $\mu \mathrm{m}$ (the electrode thickness) down to 10 nm (the cracks in active particles). As such, the model geometry depicted in Figure \ref{fig:sketch_full} should still be interpreted as an ``effective geometry''. Despite this caveat, the microscale model is still an extremely useful tool and constitutes the starting point from which to derive the other models. Examples of its use to model real battery behaviour are to be found in \cite{Less2012,Lu2020,Shodiev2020} and the commercial battery software BEST (Battery and Electrochemistry Simulation Tool) \cite{BEST}.

Before presenting the equations for the microscale model, we discuss the transport laws used to formulate the microscale model and all subsequent electrochemical models in this article. Further details on the derivation of the model can be found in the handbooks of battery modelling \cite{Newman2004,Plett2015} and in the review \cite{Richardson2021}. In the electrodes, the model equations ensure that we have conservation of intercalated lithium and electrons. The transport of lithium in the electrode particles is complex and often involves multiple phase-transitions. Despite this, the usual assumption is that transport can be modelled by a diffusion equation, albeit one in which the diffusion coefficient depends on the lithium concentration, and is thereby capable of capturing some of the important features of the phase transition behaviour. Electron transport through the matrix of electrode particles and conductive binder is assumed to be driven by Ohm's law.

The electrolyte, which fills the pores of the separator and the electrodes (comprised of electrode particles, binder and conductive additives), acts to transport both lithium ions and counterions across the cell. Since ions are charged particles there is strong coupling between the lithium ion and charge flows which, in turn, means that ion and current fluxes in the electrolyte cannot be defined independently, as they are in the electrode particles. Various different theories are used to describe charge transport in the electrolyte. The most commonly used are the ``dilute electrolyte theory'', which is based on the Nernst-Planck equations and is only really applicable for very dilute electrolytes, and the more involved ``concentrated electrolyte theory'', which is based upon the Stefan-Maxwell equations and works well at the moderate ion concentrations encountered in real batteries. Both models are presented in \cite{Newman2004} and compared in \cite{Richardson2021}. Since most battery electrolytes are too concentrated to be realistically modelled by the Nernst-Planck theory, here we consider only the concentrated electrolyte model.

Lithium ion exchange between the electrode particles and the electrolyte is typically described by a Butler-Volmer equation. The Butler-Volmer relation, although essentially \textit{ad hoc}, is widely-used and shows good agreement to experiments \cite{Dickinson2020}. Moreover, even though it is applied in the form of a boundary (or interface) condition in battery models, it is being used to capture the processes that occur both at the electrolyte/electrode interface and those that occur in its immediate vicinity (within a few Debye lengths), where charge neutrality does not apply. There are two main quantities of interest in the Butler-Volmer equation. The first is the overpotential, which is difference in the lithium ion electrochemical potential across the interface (and is therefore zero at equilibrium) while the second is the exchange current density, which is the current density passing backwards and forwards between the active material and the electrolyte in equilibrium. The overpotential depends both on the electric potentials, in the electrode and electrolyte, and on the equilibrium potential which, in turn, depends on the lithium ion concentration in the particles. The exchange current density depends on the local lithium ion concentrations in the electrolyte and at the surfaces of the particles. The four variables that influence the ion exchange reaction, between electrode and electrolyte, are thus the electric potentials and lithium ion concentrations on either side of the interface.

\paragraph{Electrode equations} Having outlined the physics that underpin mass and charge transport in a lithium-ion cell, we are now in a position to formulate the model. The geometry over which the model is solved is shown in Figure \ref{fig:sketch_full}. Lithium and charge conservation within the porous electrode matrix ($\kin{n,p}$ for negative and positive electrode, respectively) are:
\begin{subequations}\label{eq:microscale_electrode}
\begin{align}
    \pdv{c_k}{t} + \nabla \cdot \tilde{\vb*{N}}_k &= 0, & \text{ in } \vb*{x} \in \Omega_k^\micro,\\
    \nabla \cdot \tilde{\vb*{i}}_k &= 0, & \text{ in } \vb*{x} \in \Omega_k^\micro,
\end{align}
where $\Omega_k^\micro$ is the domain of the solid part (including particles and binder) of the electrode. The constitutive equations that specify the lithium ion flux $\tilde{\vb*{N}}_k$ and current density $\tilde{\vb*{i}}_k$ in terms of the gradients in the lithium concentration $c_k$ and the electrical potential $\phi_k$ are, respectively,
\begin{align}
    \tilde{\vb*{N}}_k &= - \tilde{D}_k(c_k,\vb*{x}) \nabla c_k,& \text{ in } \vb*{x} \in \Omega_k^\micro,\\
    \tilde{\vb*{i}}_k &= - \tilde{\sigma}_k(\vb*{x}) \nabla \phi_k, & \text{ in } \vb*{x} \in \Omega_k^\micro.
\end{align}
\end{subequations}
Since the domain $\Omega_k^\micro$ contains both electrode particles and binder, the diffusivity and conductivity are both spatially dependent. In particular, given that lithium ions can only diffuse in the active material of the electrode particles, the lithium diffusivity $\tilde{D}_k=0$ in those parts of the domain occupied by binder. The lithium diffusivity is also typically a strong function of concentration (see e.g. \cite{Ecker2015i}). The electronic conductivity, $\tilde{\sigma}_k(\vb*{x})$, typically varies strongly over the microscale between those parts of the domain occupied by binder and those occupied by electrode particles \cite{Cheng2019}. 

The tildes used in (\ref{eq:microscale_electrode}a)-(\ref{eq:microscale_electrode}d) are to distinguish microscale fluxes and parameters from the averaged macroscale fluxes and parameters (i.e. after homogenisation), which are used in Section \ref{sec:homogenisation} onward. This distinction, even though subtle, is crucial from the modelling point of view. For the microscale model, we have a resolved geometry that discerns the porous structure of the electrodes and therefore both the fluxes and parameters can be defined throughout the relevant parts of the structure. Post homogenisation, we obtain a homogenised geometry that does not resolve the porous microstructure, but the macroscale fluxes and parameters are averaged to account for the effects of the microstructure, as discussed in Section \ref{sec:homogenisation}.

\paragraph{Electrolyte equations} The electrolyte fills the region $\Omega_\mre^\micro$, which corresponds to all the pores in the electrodes and separator. On defining the concentration of positive lithium ions as $c_{\mre +}$ and that of the negative counterions as $c_{\mre -}$, charge neutrality allows us to write $c_{\mre +}=c_{\mre -}=c_{\mre}$. Then, conservation laws for positive (lithium) and negative ions, respectively, take the form:
\begin{subequations}\label{eq:cons_ions_microscale}
\begin{align}
    \pdv{c_{\mre}}{t} + \nabla \cdot \tilde{\vb*{N}}_{\mre +} &= 0, & \text{ in } \vb*{x} \in \Omega_\mre^\micro,\label{eq:cons_ions_microscale1}\\
    \pdv{c_{\mre}}{t} + \nabla \cdot \tilde{\vb*{N}}_{\mre -} &= 0, & \text{ in } \vb*{x} \in \Omega_\mre^\micro. \label{eq:cons_ions_microscale2}
\end{align}
Here $\tilde{\vb*{N}}_{\mre +}$ and $\tilde{\vb*{N}}_{\mre -}$ are the lithium ion flux and the flux of the negatively charged counterions, respectively. Constitutive laws for these two ion fluxes are derived from the Stefan-Maxwell equations. In the case of a moderately concentrated electrolyte, in which the solvent is at much higher concentration than the ion species (see \cite{Newman2004,Richardson2021} for details), these yield the following constitutive laws for the ion fluxes:
\begin{align} \label{constitutive1}
    \tilde{\vb*{N}}_{\mre +} &= - \tilde{D}_\mre \nabla c_\mre + \frac{t^+}{F} \tilde{\vb*{i}}_\mre,\\
    \label{constitutive2} 
    \tilde{\vb*{N}}_{\mre -} &= - \tilde{D}_\mre \nabla c_\mre - \frac{1 - t^+}{F} \tilde{\vb*{i}}_\mre,
\end{align}
\end{subequations}
in which $\tilde D_\mre$ is the ionic diffusivity, $t^+$ is the transference number (i.e. the fraction of the current carried by positive ions), $F$ is the Faraday constant and $\tilde{\vb*{i}}_\mre$ is the current density. In general, both $\tilde D_\mre$ and $t^+$ depend on ion concentration $c_\mre$.

Subtracting \eqref{constitutive2} from \eqref{constitutive1} yields the expression for the current density in the electrolyte
\begin{equation}
\tilde{\vb*{i}}_\mre = F (\tilde{\vb*{N}}_{\mre +} - \tilde{\vb*{N}}_{\mre -}), \label{current-eq}
\end{equation}
and then, by subtracting \eqref{eq:cons_ions_microscale2} from \eqref{eq:cons_ions_microscale1} and utilising \eqref{current-eq}, we obtain the current conservation law
\begin{equation}\label{eq:charge-cons}
    \nabla \cdot \tilde{\vb*{i}}_\mre = 0. 
\end{equation}
An expression for $\tilde{\vb*{i}}_\mre$, in terms of $c_\mre$ and $\phi_\mre$, is obtained from the Stefan-Maxwell equations by balancing the forces arising from the gradients in the electrochemical potentials of both ion species with the drag between the two ion species. By following this procedure, it can be shown (see \cite{Newman2004,Plett2015,Richardson2021} for details) that the constitutive law for the current density is
\begin{equation}
\tilde{\vb*{i}}_\mre = - \tilde{\sigma}_\mre \left(\nabla \phi_\mre - \frac{2}{F} (1 - t^+) \nabla \mu_\mre \right). \label{current}
\end{equation}
Here $\phi_\mre$ is the electric potential of the electrolyte measured with respect to a lithium electrode and $\mu_\mre(c_\mre)$ is the chemical potential of the electrolyte. The former can be thought of as the electrochemical potential of the lithium ions, $\bar{\mu}_{\mre +}$, divided Faraday's constant, i.e. $\phi_\mre=\bar{\mu}_{\mre +}/F$, while the latter is defined in terms of the chemical potentials of the lithium ions and the negative counterions ($\mu_{\mre +}$ and $\mu_{\mre -}$, respectively) by the relation $\mu_\mre=\frac{1}{2}(\mu_{\mre +}+\mu_{\mre -})$.

In order to arrive at a closed model, we require two conservation equations (out of \eqref{eq:cons_ions_microscale1}, \eqref{eq:cons_ions_microscale2} and \eqref{eq:charge-cons}) and the two corresponding constitutive laws (out of \eqref{constitutive1}, \eqref{constitutive2} and \eqref{current}). Probably the most common choice, and the one we adopt here, is to take the equations for the lithium ions, \eqref{eq:cons_ions_microscale1} and \eqref{constitutive1}, and for the current, \eqref{eq:charge-cons} and \eqref{current}. This choice yields the following closed set of electrolyte equations:
\begin{subequations}\label{eq:microscale_electrolyte}
\begin{align}
    \pdv{c_\mre}{t} + \nabla \cdot \tilde{\vb*{N}}_{\mre +} &= 0, & \text{ in } \vb*{x} \in \Omega_\mre^\micro,\\
    \nabla \cdot \tilde{\vb*{i}}_\mre &= 0, & \text{ in } \vb*{x} \in \Omega_\mre^\micro,
\end{align}
with the fluxes
\begin{align}
\tilde{\vb*{N}}_{\mre +} &= - \tilde{D}_\mre \nabla c_\mre + \frac{t^+}{F} \tilde{\vb*{i}}_\mre,\\
\tilde{\vb*{i}}_\mre &= - \tilde{\sigma}_\mre \left(\nabla \phi_\mre - \frac{2}{F} (1 - t^+) \dv{\mu_\mre}{c_{\mre}} \nabla c_\mre \right).
\end{align}
\end{subequations}
From a mathematical perspective it is sometimes more convenient to works in terms of the counterion concentration and flux (as opposed to those of the lithium ions) as this makes implementation of a conservative numerical scheme in the electrolyte somewhat easier.

Here, the model variables are $c_\mre$ the lithium ion concentration, $\phi_\mre$ the electrolyte potential (measured with respect to a lithium electrode), $\tilde{\vb*{N}}_{\mre +}$ the flux of lithium ions, and $\tilde{\vb*{i}}_\mre$ the current density. Electrolyte diffusivity $\tilde{D}_\mre$, ionic conductivity $\tilde{\sigma}_\mre$ and electrolyte chemical potential $\mu_{\mre}$ are all usually considered to be functions of the ion concentration $c_{\mre}$ and the temperature $T$, and must be fitted to data (see, for example, \cite{Ecker2015i,Landesfeind2019}), though it is fairly standard to assume that electrolyte is ideal such that the chemical potential is given by $\mu_{\mre} = R T \log c_{\mre}$. The transference number $t^+$ is in general a function of $c_\mre$ and $T$ but is often found to be close to a constant.

\paragraph{Boundary and initial conditions} To close the model boundary and initial conditions need to be enforced. Internal boundary conditions are imposed at the interface between the electrolyte and the electrode particles, while external boundary conditions are imposed on boundaries with other components of the battery, such as the current collectors or the separator.

Internal boundary conditions that encapsulate the effects of the lithium intercalation reactions on the surfaces of the electrode particles can be written in the form
\begin{subequations}\label{eq:microscale_internal_bc}
\begin{align}
     F \tilde{\vb*{N}}_k \cdot \vb*{n}_k &= \tilde{\vb*{i}}_k \cdot \vb*{n}_k = \tilde{j}_k, & \text{ at } \vb*{x} \in \partial \Omega_k^\mathrm{in},\\
     F \tilde{\vb*{N}}_\mre \cdot \vb*{n}_k &= \tilde{\vb*{i}}_\mre \cdot \vb*{n}_k = -\tilde{j}_k, & \text{ at } \vb*{x} \in \partial \Omega_\mre^\mathrm{in},
\end{align}
\end{subequations}
for $\kin{n,p}$, where $\tilde{j}_k$ is the reaction current density flowing through the surfaces of the particles and $\vb*{n}_k$ is the unit normal vector (pointing outwards) from the boundary  $\partial \Omega_k^\micro$ of the particle/binder domain. These conditions represent conservation of charge and lithium ions on the interface (i.e. what leaves/enters the electrode must enter/leave the electrolyte) and the fact that the current flowing into the electrolyte is carried entirely by lithium ions.

The reaction current density, $\tilde{j}_k$, is determined by the Butler-Volmer relation
\begin{subequations}\label{eq:microscale_reaction}
\begin{align}
    \tilde{j}_k &= j_{k0} \sinh \left( \frac{F}{2 R T} \eta_k \right),\\
    j_{k0} &= F K_k \sqrt{\frac{c_\mre}{c_{\mre 0}} \frac{c_k}{c_k^{\max}} \left(1 - \frac{c_k}{c_k^{\max}} \right) },\label{eq:exchange_current_dens_BV}\\
    \eta_k &= \phi_k - \phi_\mre - U_k(c_k),
\end{align}
\end{subequations}
for $\kin{n,p}$. Here $\tilde{j}_k$ is the de-intercalation current density, $j_{k0}$ is the exchange current density and $\eta_k$ is the overpotential, where $K_k$ is the reaction constant, $c_{\mre 0}$ is the initial ion concentration in the electrolyte, $c_k^{\max}$ is the maximum lithium concentration in the particles and $U_k$ is the open-circuit potential. Note that the reaction constant, $K_k$, has units of mol s$^{-1}$ m$^{-2}$ here, but in the literature it is common to find equivalent, yet different, formulations of \eqref{eq:exchange_current_dens_BV}, where the counterpart constant has other units. The reaction constant $K_k$ is zero on those parts of $\partial \Omega_k^\micro$ where the interface is between the electrolyte and the binder.

Notably, the reaction rate is zero when $\eta_k=0$ and lithium ions in the electrolyte are in electrochemical equilibrium with lithium ions in the active material. The key new function that appears in this equation and encapsulates information about the chemical energy associated with intercalating lithium ions into the active material of the electrode particles, is the open-circuit potential, $U_k(c_k)$. This may be measured experimentally by placing a lithium electrode in the electrolyte and measuring the potential difference between this and the active material at equilibrium. Note that here we have assumed a symmetric Butler-Volmer reaction, but other authors have examined non-symmetric expressions (see \cite{Newman2004,Plett2015,Richardson2021} for details). All the results in this paper can be readily extended for non-symmetric reactions.

Different external boundary conditions need to be imposed on the electrode and the electrolyte. For the electrode a distinction needs to be drawn between the electrode-current collector boundary, $\partial \Omega_k^\mathrm{cc}$, and the rest of the outer boundary, $\partial \Omega_k^\mathrm{out}$ , including the electrode-separator interface (see Figure \ref{fig:sketch_full}). Current can enter the electrode through the current collector boundary in the form of conduction electrons, while the rest of the outer boundary is impermeable to charge (in terms of both lithium and electron flux). There are multiple boundary conditions that can be imposed at the current collector. A natural way to understand the boundary conditions on the electrode is to prescribe a potential difference between the two current collectors (which are assumed to be equipotential), which will induce a current into the battery. For the electrolyte, on the other hand, the whole outer boundary, $\partial \Omega_\mre^\mathrm{out}$, is impermeable. Thus, the boundary conditions can be written as
\begin{subequations}\label{eq:microscale_external_bc}
\begin{align}
    \tilde{\vb*{N}}_k \cdot \vb*{n}_k &= 0, & & & \text{ in } \vb*{x} \in \partial \Omega_k^{\mathrm{cc}},\\
    \tilde{\vb*{N}}_k \cdot \vb*{n}_k &= 0, & \tilde{\vb*{i}}_k \cdot \vb*{n}_k &= 0, & \text{ in } \vb*{x} \in \partial \Omega_k^{\mathrm{out}},\\
    \tilde{\vb*{N}}_\mre \cdot \vb*{n}_\mre &= 0, & \tilde{\vb*{i}}_\mre \cdot \vb*{n}_\mre &= 0, & \text{ in } \vb*{x} \in \partial \Omega_\mre^{\mathrm{out}},
\end{align}
\end{subequations}
for $\kin{n,p}$. We prescribe a given potential difference $V(t)$ between the current collectors by setting
\begin{align}\label{eq:microscale_potential_bc}
    \phi_\mrn &= 0, & \text{ at } \vb*{x} \in \partial \Omega_\mrn^\mathrm{cc},\\
    \phi_\mrp &= V(t), & \text{ at } \vb*{x} \in \partial \Omega_\mrp^\mathrm{cc}.
\end{align}
Note that we set the reference potential in the negative current collector, but this is an arbitrary choice. Quite often we need to prescribe a total applied current to the battery $I(t)$ rather than a voltage. Then, the potential difference $V(t)$ is unknown and must be adjusted to satisfy the condition
\begin{equation}
    I(t) = \int_{\partial \Omega_k^{\mathrm{cc}}} \tilde{\vb*{i}}_k \cdot \vb*{n}_k \dd \vb*{x},
\end{equation}
for $\kin{n,p}$.

The model is closed by posing initial conditions for the electrode and electrolyte concentrations:
\begin{align}\label{eq:microscale_ic}
    c_k &= c_{k0}, & c_\mre &= c_{\mre 0}, & \text{ at } t = 0.
\end{align}

\paragraph{Summary of the microscale model} Equations \eqref{eq:microscale_electrode} and \eqref{eq:microscale_electrolyte}, together with the boundary and initial conditions \eqref{eq:microscale_internal_bc}-\eqref{eq:microscale_ic}, comprise the microscale model. The model equations can be sub-divided into conservation laws (as defined in (\ref{eq:microscale_electrode}a), (\ref{eq:microscale_electrode}b), (\ref{eq:microscale_electrolyte}a) and (\ref{eq:microscale_electrolyte}b)) and transport laws (as defined in (\ref{eq:microscale_electrode}c), (\ref{eq:microscale_electrode}d), (\ref{eq:microscale_electrolyte}c) and (\ref{eq:microscale_electrolyte}d)). Whilst the boundary conditions (given by \eqref{eq:microscale_internal_bc} and \eqref{eq:microscale_external_bc}) can be divided into external and internal conditions, in which the latter (see \eqref{eq:microscale_internal_bc}) captures the intercalation reaction. This reaction is typically described by Butler-Volmer kinetics (see \eqref{eq:microscale_reaction}). We note that there is a clear correspondence between the physical laws described at the beginning of Section \ref{sec:microscale} and the model equations and boundary conditions.

\subsubsection{Limitations and remarks about the microscale model}
In defining the microscale model a number of choices were made about how to model the various physical phenomena that determine battery performance. These were broadly in line with the usual approach adopted in the literature, nevertheless it can be argued that different modelling choices could improve the accuracy of the microscale model. In what follows, we address some of these alternative approaches and remark upon some of the technical nuances of the model.

Here lithium transport in the electrode particles is modelled by nonlinear diffusion equations, (\ref{eq:microscale_electrode}a) and (\ref{eq:microscale_electrode}c). Despite this, it is common in the literature to model this process using a linear (Fickian) diffusion equation (i.e. one with constant diffusion coefficient), which is quite an unrealistic assumption (see, for example, the differences between \cite{Chen2020} and \cite{ORegan2022}). Moreover, it has been argued by Bazant and others, e.g. \cite{Singh2008,Thomas-Alyea2017,Zeng2013,Zeng2014}, that lithium ion transport in phase change electrode materials is better described by a phase-field model than by a nonlinear diffusion equation; this extension to the model is discussed in more detail in Section \ref{sec:extensions}.

A key point, which is often overlooked in the literature, is that the electrolyte potential $\phi_\mre(x)$, which is used in the model of the electrolyte, is not a true electrical potential. Rather, it is the electrical potential that would be measured by inserting a lithium reference electrode into the electrolyte. As discussed in \cite{Bizeray2016,Ranom2014,Richardson2021}, this results in a relation between the true electrical potential in the electrolyte, $\varphi$, and the potential measured with respect to a lithium reference electrode, $\phi_\mre$, which reads
\begin{equation}
\phi_\mre=\varphi-\frac{RT}{F}\log(a_\mrp) + \text{constant},
\end{equation}
where $a_\mrp$ is the activity of the lithium ions in the electrolyte. As such, $\phi_\mre$ is closely related to the \textit{electrochemical} potential of lithium ions in the electrolyte, $\mu_{\mre +}$, via the equation $\mu_{\mre +} = F \phi_\mre + \text{constant}$. This, as pointed out in \cite{Bizeray2016,Ranom2014,Richardson2021}, has led to numerous discrepancies in the literature. Another point to highlight is that the electrochemical transport theory underlying the concentrated electrolyte equations has been further developed in \cite{Goyal2017,Goyal2021,Liu2014}; in particular, streamlining thermodynamic foundations of the theory to account for temperature, pressure and stress effects, and clarifying the relationship between the chemical and electrical potentials. Finally, we want to highlight that, quite often, the gradient of the electrochemical potential is written in terms of $f_{\pm}$, the mean molal activity coefficient of the salt. From \cite{Plett2015} it can be seen that for a moderately concentrated electrolyte, the gradient of the chemical potential, $\mu_\mre$, can be replaced by the expression
\begin{equation}
\nabla \mu_\mre = RT \left( 1 + \pdv{\log f_\pm}{\log c_\mre} \right) \nabla \log c_\mre, \label{drivel}
\end{equation}
where $\pdv{\log f_\pm}{\log c_\mre}$ is usually determined from experimental data.

\subsection{Homogenised model}
\label{sec:homogenisation}

\begin{figure}
    \centering
    \includegraphics[width=\linewidth]{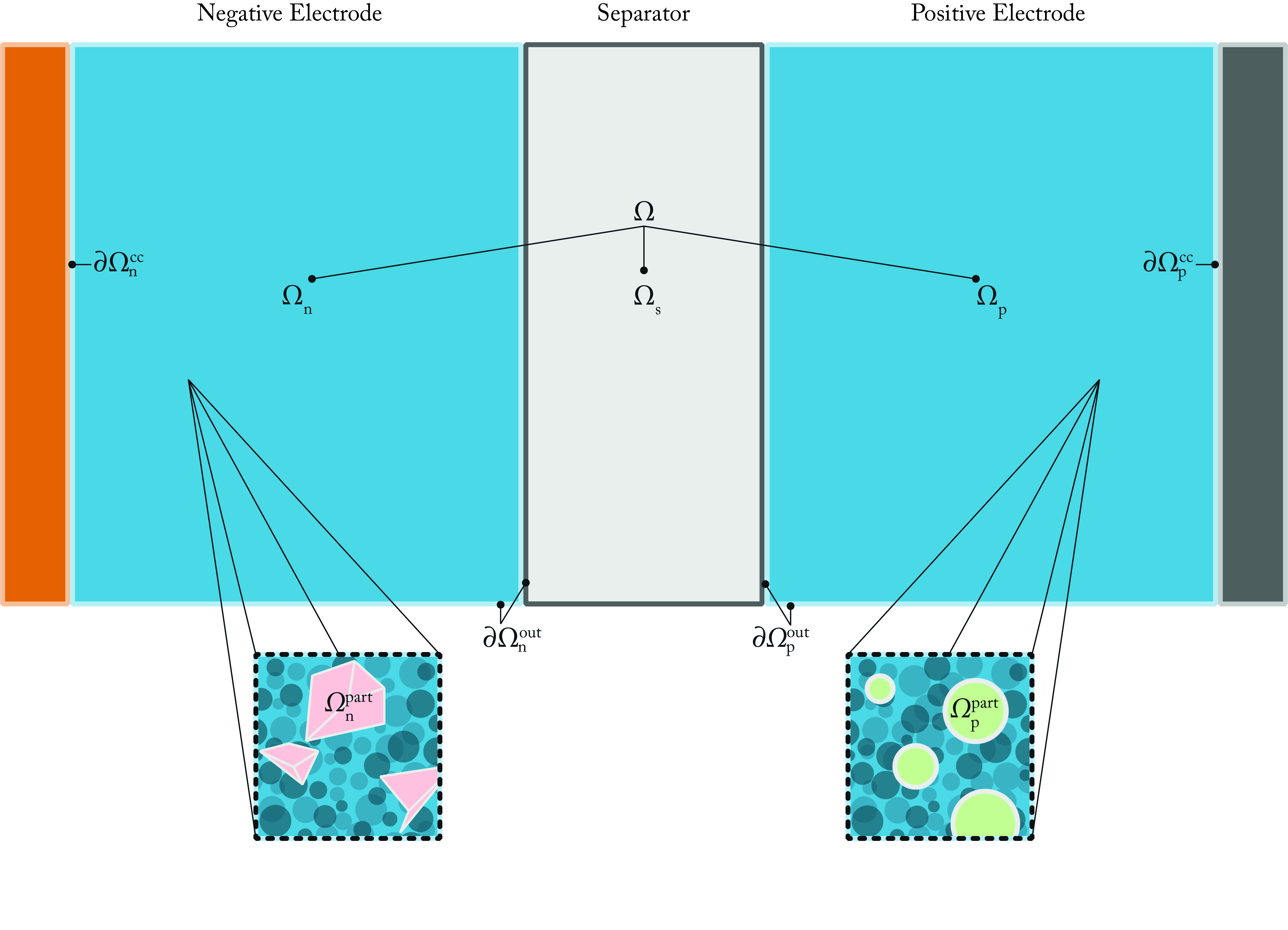}
    \caption{Sketch of the geometry of the homogenised model. This is a multiscale model, so the macroscale domains (electrodes $\Omega_\mrn$ and $\Omega_\mrp$, and separator $\Omega_\mrs$) are treated as a continuous material. At each point in the electrodes, there is a representative microscale domain (electrode particles $\Omega_\mrn^\text{part}$ and $\Omega_\mrp^\text{part}$). Note that the domain, $\Omega$, is defined as $\Omega = \Omega_\mrn \cup \Omega_\mrs \cup \Omega_\mrp$.}
    \label{fig:sketch_homogenised}
\end{figure}

In practice, it is extremely computationally expensive to solve the model described in Section \ref{sec:microscale} because of the complexity of the geometry. In particular, the electrode equations \eqref{eq:microscale_electrode} need to be solved throughout the porous matrix, which is composed of electrode particles, binder and conductive additives (all of which have very different properties), whilst the electrolyte equations \eqref{eq:microscale_electrolyte} are solved in the highly tortuous pore space of this matrix. However, solutions to the microscale model suggest that many of its variables change significantly only over length scales much larger than the microstructure. This suggests that a much simpler, homogenised model can be used to accurately approximate the full problem. Some examples of the applications of homogenised models to study real batteries are \cite{Kashkooli2017,Kashkooli2016,Kim2018}.

The key idea of homogenised models is to treat the porous material and its pore space as a continuum, by modifying the equations to account for the effects of the microstructure. There are three main changes that occur to these equations after they are homogenised. First, the fluxes in the equations are re-defined as averaged fluxes in the new homogenised geometry, as is the differential operator, $\nabla$. Second, \textit{effective} transport parameters are used to account for the tortuosity effects of the highly convoluted porous geometry in the transport equations. Third, the reactions that appeared as boundary conditions at the electrode-electrolyte interface, \eqref{eq:microscale_internal_bc}-\eqref{eq:microscale_reaction}, now appear as source terms in the bulk equations, since the electrode-electrolyte interface is not explicitly captured in the homogenised geometry. The main advantage of the homogenised equations is that they depend only upon a macroscopic spatial variable and are solved in a much simpler ``homogeneous'' geometry, as shown in Figure \ref{fig:sketch_homogenised}. This allows a much simpler mesh and a much coarser spatial discretisation to be used compared to the microscale model.

However, the homogenised battery model (unlike the porous medium flow equation, see \cite{Whitaker1986}) retains a microscale variable in order to model lithium transport within the electrode particles. This is caused by two main reasons. First, lithium-ion motion within the active materials that form the particles is extremely slow, so that significant lithium-ion concentration gradients form on the microscale in order to drive the requisite fluxes. And second, because lithium-ions cannot flow directly between the discrete microscopic electrode particles. Therefore, when the microscale model is homogenised the particle equations do not upscale and only the electrolyte equations (\ref{eq:microscale_electrolyte}) and the current flow equations in the matrix (\ref{eq:microscale_electrode}b) and (\ref{eq:microscale_electrode}d) do (see \cite{Hunt2020}). These macroscopic equations couple to a series of microscale problems, parameterised by the macroscopic variable $\vb*{X}$, describing lithium-ion transport within the particles at representative points in the electrodes.

From this coupling between the macro- and microscale equations, we get the classic multiscale formulation of the porous electrode model. In fact, the widely used Doyle-Fuller-Newman model is conceptually a homogenised model, even though in \cite{Doyle1993,Fuller1994a,Fuller1994} it was originally justified on an \textit{ad hoc} basis, rather than being formally derived. However, asymptotic homogenisation offers us a tool to formally derive the homogenised model (including the equations and the effective parameters) from the full equations at the microscale in a systematic manner \cite{Arunachalam2015,Ciucci2011,Hennessy2020,Hunt2020,Richardson2012}. Asymptotic homogenisation is a well-established mathematical technique and the technical details can be found in many books (e.g. \cite{Bensoussan2011,Pavliotis2008}). The key idea of this method is that, if the model has two (or more) very distinct length scales (such as particles and whole electrodes in batteries), they can be treated as independent in terms of the spatial mathematical operators. This scale separation leads, through a systematic procedure, to the homogenised model. Homogenisation might appear analogous to the volume averaging method, \cite{Plett2015,Whitaker1999} however, the latter requires an \textit{ad hoc} closure condition, usually empirically defined, in order to complete the problem. As discussed in \cite{Davit2013}, in which a comparison is made between the two methodologies, the main differences are in the assumptions and formalisms of the two methods, rather than in the results they yield.

\begin{figure}
    \centering
    \includegraphics[width=\linewidth]{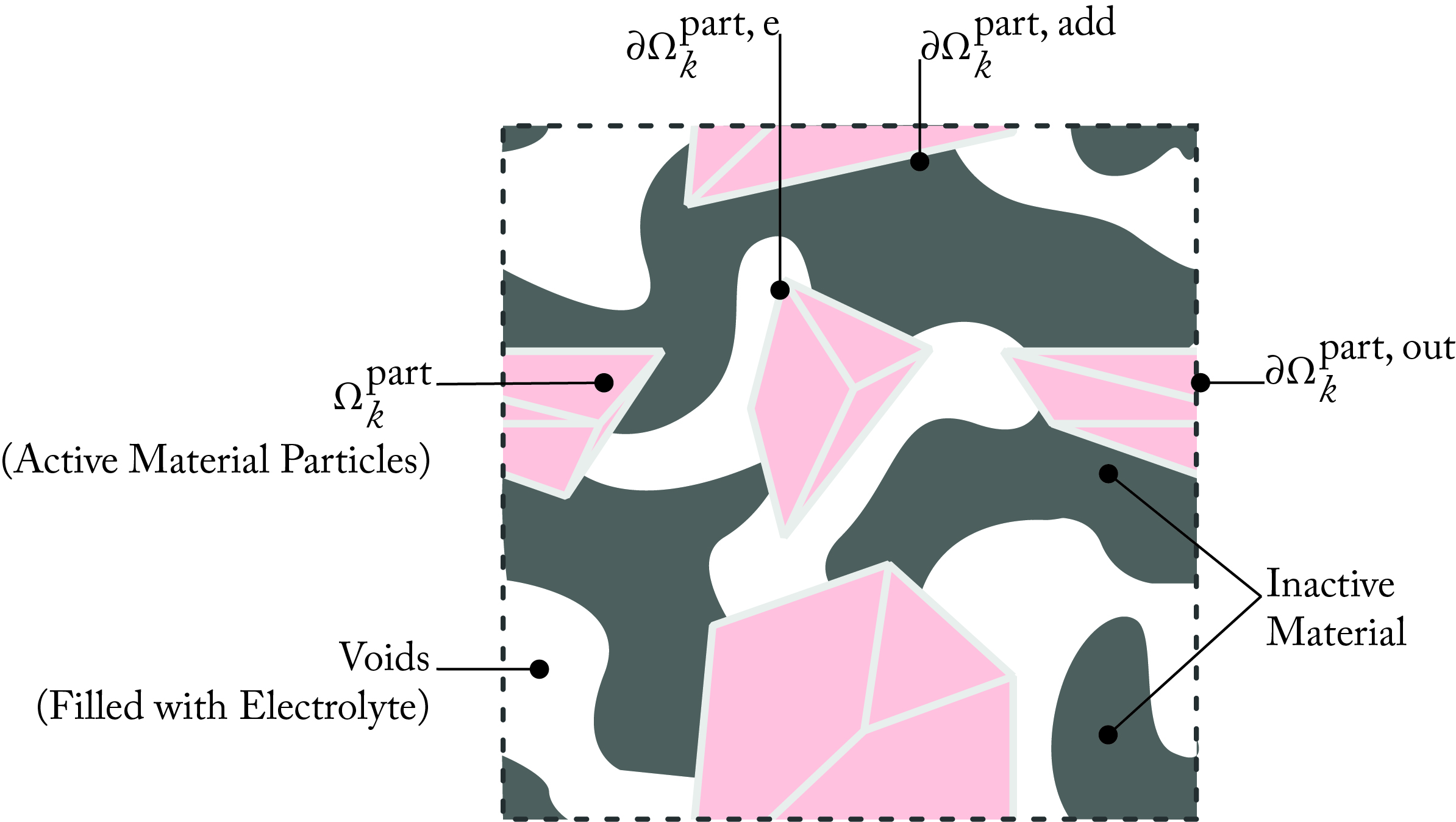}
    \caption{Sketch of the microstructure of the homogenised model. The solid matrix is composed of the particles $\Omega_k^\mathrm{part}$ and the inactive material where lithium cannot intercalate. The boundary of the particles is divided into the part in contact with the electrolyte $\partial \Omega_{k}^\mathrm{part,e}$, the part in contact with the inactive additives $\partial \Omega_{k}^\mathrm{part,add}$, and the intersection with the representative microstructure boundary $\partial \Omega_{k}^\mathrm{part,out}$. The geometry is periodic, which means that if we tessellate the space with the representative volume element, each subdomain would connect with itself across the boundary. Note that this is a two-dimensional, schematic representation and, in order for the solid and electrolyte domains to be connected, we would require a three-dimensional geometry.}
    \label{fig:sketch_micro_homogenisation}
\end{figure}

A statement of this model is given below for the geometry depicted in Figures \ref{fig:sketch_homogenised} and \ref{fig:sketch_micro_homogenisation}. The variable $\vb*{x}$ is the macroscale spatial variables, used to measure distance across the electrodes and electrolyte, while the variable $\vb*{X}$ is the microscale spatial variable, used to measure distances in individual electrode particles).

\paragraph{Particle equation} Then, lithium concentration within representative electrode particles is computed on a series of microscale domains, at each point in the macroscopic space, and is governed by
\begin{subequations}\label{eq:homogenised_particle}
\begin{align}
    \pdv{c_k}{t} + \nabla_{X} \cdot \tilde{\vb*{N}}_k &= 0, & \text{ in } \vb*{X} \in \Omega_{k}^\mathrm{part},\\
    F \tilde{\vb*{N}}_k \cdot \vb*{n}_{k} &= \tilde{j}_k, & \text{ at } \vb*{X} \in \partial \Omega_{k}^\mathrm{part,e},\\
    F \tilde{\vb*{N}}_k \cdot \vb*{n}_{k} &= 0, & \text{ at } \vb*{X} \in  \partial \Omega_{k}^\mathrm{part,add},\\
    \text{periodic } & \text{conditions},  & \text{ at } \vb*{X} \in \partial \Omega_{k}^\mathrm{part,out},\\
    c_k &= c_{k0}, & \text{ at } t = 0,
\end{align}
with
\begin{equation}
    \tilde{\vb*{N}}_k = - \tilde{D}_k  \nabla_X c_k.
\end{equation}
\end{subequations}

To be clear, when we say ``periodic conditions'', we mean that the concentrations and fluxes on the boundaries of the domain must match those on neighbouring boundary segments, when the representative volume element tesselates the space. The concentration of intercalated lithium is denoted by $c_k(\vb*{X},\vb*{x},t)$, and the lithium flux by $\tilde{\vb*{N}}_k$, where subscript $\kin{n,p}$ represents negative and positive electrode particles, respectively. The other symbols have been introduced previously, but a nomenclature table is available in \ref{sec:nomenclature}.

The first difference between the homogenised model and the microscale model, presented in Section \ref{sec:microscale}, is that the former assumes that a representative microstructure repeats periodically (though the model could be extended to account for a slowly varying, microscale geometry). The second difference is that, in the homogenised model, we can just consider the diffusion in the active material particles, as opposed to considering the whole porous matrix, as lithium cannot diffuse into the binder and other additives, so the domain is $\Omega_{k}^\mathrm{part}$. The boundary of active material particles is split into three subdomains: the part in contact with the electrolyte $\partial \Omega_{k}^\mathrm{part,e}$, the part in contact with the inactive additives $\partial \Omega_{k}^\mathrm{part,add}$, and the intersection with the representative microstructure boundary $\partial \Omega_{k}^\mathrm{part,out}$. The domains are illustrated in Figure \ref{fig:sketch_micro_homogenisation}.

\paragraph{Electrode equation} The homogenised macroscale equations describing electron conduction in the solid matrices of the two electrodes are
\begin{subequations}\label{eq:homogenised_electrode}
\begin{align}
    \nabla \cdot \vb*{i}_k &= - b_k j_k, & \text{ in } \vb*{x} \in \Omega_{k},\\
    \vb*{i}_k \cdot \vb*{n}_k &= i_\mathrm{app}, & \text{ at } \vb*{x} \in \partial \Omega_k^\mathrm{cc},\\
    \vb*{i}_k \cdot \vb*{n}_k &= 0, & \text{ at } \vb*{x} \in \partial \Omega_k^\mathrm{out},
\end{align}
with
\begin{equation}
    \vb*{i}_k = - \sigma_k \nabla \phi_k,
\end{equation}
\end{subequations}
for $\kin{n,p}$. Here the electric potential in the electrode is denoted by $\phi_k$; the current density averaged over the porous electrode structure by $\vb*{i}_k$; the surface reaction current averaged over the surface of the porous matrix in contact with the electrolyte by $j_k$; the surface area (per unit volume) of porous matrix in contact with the electrolyte by $b_k$; the effective electronic conductivity by $\sigma_k$,  and the applied current density averaged over the current collectors by $i_\mathrm{app}$. Note that here (\ref{eq:homogenised_electrode}b) replaced the potential condition for the microscale model \eqref{eq:microscale_potential_bc} but we could use the latter too.

Because this is a homogenised model, there are some significant differences with the microscale model, \eqref{eq:microscale_electrode}. First, the electrode current density, $\vb*{i}_k$, is now defined as a current density in the homogenised medium rather than as a current density in the porous material. Therefore, the electronic conductivity, $\sigma_k$, is now an effective parameter accounting for the heterogeneous porous structure. The reaction current, $j_k$, is defined as the average over the electrode-electrolyte interface
\begin{equation}
    j_k = \frac{1}{| \partial \Omega_k^\mathrm{part,e} |} \int_{\partial \Omega_k^\mathrm{part,e}} \tilde{j}_k \dd \vb*{X}.
\end{equation}
To distinguish between the parameters, fluxes and currents at the microscale from the homogenised ones, we use tildes for the first and drop the tildes for the latter.  

\paragraph{Electrolyte equations}
The homogenised macroscopic electrolyte equations are
\begin{subequations}\label{eq:homogenised_electrolyte}
\begin{align}
    \varepsilon(\vb*{x}) \pdv{c_\mre}{t} + \nabla \cdot \vb*{N}_\mre &= \frac{b(\vb*{x}) j(\vb*{x},t)}{F} & \text { in } \vb*{x} \in \Omega,\\
    \nabla \cdot \vb*{i}_\mre &= b(\vb*{x}) j(\vb*{x},t) & \text { in } \vb*{x} \in \Omega,\\
    \vb*{N}_\mre \cdot \vb*{n}_\mre &= 0, & \text{ at } \vb*{x} \in \partial \Omega,\\
    \vb*{i}_\mre \cdot \vb*{n}_\mre &= 0, & \text{ at } \vb*{x} \in \partial \Omega,\\
    c_\mre &= c_{\mre 0}, & \text{ at } t = 0,
\end{align}
with:
\begin{align}
    \vb*{N}_\mre &= - D_\mre \nabla c_\mre + \frac{t^+}{F} \vb*{i}_\mre,\\
    \vb*{i}_\mre &= - \sigma_\mre \left(\nabla \phi_\mre - \frac{2}{F} (1 - t^+) \dv{\mu_\mre}{c_\mre} \nabla c_\mre \right).
\end{align}
where the porosity, active surface area per unit volume and reaction current density are defined, respectively, as
\begin{align}
    \varepsilon (\vb*{x}) &= \begin{cases}
    \varepsilon_\mrn, & \text{ if } \vb*{x} \in \Omega_\mrn,\\ 
    \varepsilon_\mrs, & \text{ if } \vb*{x} \in \Omega_\mrs,\\ 
    \varepsilon_\mrp, & \text{ if } \vb*{x} \in \Omega_\mrp,
    \end{cases} & 
    b (\vb*{x}) &= \begin{cases}
    b_\mrn, & \text{ if } \vb*{x} \in \Omega_\mrn,\\ 
    b_\mrs, & \text{ if } \vb*{x} \in \Omega_\mrs,\\ 
    b_\mrp, & \text{ if } \vb*{x} \in \Omega_\mrp,
    \end{cases} & 
    j (\vb*{x},t) &= \begin{cases}
    j_\mrn(\vb*{x},t), & \text{ if } \vb*{x} \in \Omega_\mrn,\\ 
    0, & \text{ if } \vb*{x} \in \Omega_\mrs,\\ 
    j_\mrp(\vb*{x},t), & \text{ if } \vb*{x} \in \Omega_\mrp.
    \end{cases}
\end{align}
\end{subequations}
Here the electrolyte concentration is given by $c_\mre$, the electrolyte potential by $\phi_\mre$, the averaged molar flux by $\vb*{N}_\mre$, and the averaged current density by $\vb*{i}_\mre$. The electrolyte parameters are the effective ionic diffusivity $D_\mre$ and the effective ionic conductivity $\sigma_\mre$, all other parameters and functions are as defined in the microscale model.

\subsubsection{Homogenised parameters}
The homogenised parameters, sometimes also called the effective parameters, can be systematically computed from the microscale parameters and the microstructure as part of the asymptotic homogenisation procedure \cite{Hunt2020,Richardson2012}. However, in many situations, this is not possible due to the lack of imaging data and therefore other techniques are used to estimate the effective homogenised parameters, as shown in Table \ref{tab:homogenised_parameters}. A detailed discussion on how to determine these parameters can be found in \cite{LeHoux2020a,Wang2022}.

\begin{table}
    \centering
    \begin{tabular}{|c|l|p{8cm}|}\hline
        $\varepsilon_k$ & Porosity & Gravimetric, imaging  \\
        $b_k$ & Active surface area per unit volume & EIS, BET, imaging\\
        $\sigma_k$ & Effective electrode conductivity & Micro modelling, EIS, ex-situ conduction experiments\\
        $D_\mre$ & Effective electrolyte diffusivity & Micro modelling, EIS, ex-situ diffusion experiments\\
        $\sigma_\mre$ & Effective electrolyte conductivity & Micro modelling, EIS, ex-situ conduction experiments\\
        $\tau$ & Tortuosity factor & As above, fitting, micro modelling, DNS, Monte-carlo, empirical\\
        \hline
    \end{tabular}
    \caption{Techniques used to characterise the homogenised parameters. For a full review, see \cite{LeHoux2020a,Wang2022}.}
    \label{tab:homogenised_parameters}
\end{table}

The porosity $\varepsilon(\vb*{x})$ is the volume fraction of the domain filled with electrolyte. Porosity may vary spatially, due to advanced electrode manufacturing techniques, and temporally, due to electrode expansion or pore clogging. However, the porosity is usually treated as a constant within each domain and can be estimated by weighing components with prior knowledge of the weight of their constituents or, more accurately, with imaging techniques.

The active surface area  $b(\vb*{x})$ is the interfacial surface area between the active material and the electrolyte, per unit volume of electrode. The interfacial surface is where the intercalation reaction takes place and thus this parameter plays an important role in modelling this reaction. Electrochemical Impedance Spectroscopy (EIS) and Brunauer–Emmett–Teller plot (BET plot)  can both be used to determine this property, as can imaging techniques.

Effective transport properties, such as diffusivity and conductivity, are always smaller than bulk transport properties and account for the fact that, in the original geometry, transport is hindered by obstacles of the non-transporting phase. Here we distinguish between the properties in electrode, ($\sigma_k$), and electrolyte, ($D_\mre$ and $\sigma_\mre$). The effective electrode conductivity needs to account for the microstructure, but also the heterogeneities in the material (i.e. different components such as active material, carbon black, binder, etc.). The effective electrolyte properties, namely the diffusivity, $D_\mre$, and conductivity, $\sigma_\mre$, account for geometrical effects only, and thus it can be written as
\begin{align}
    D_\mre &= \tilde{D}_\mre \mathcal{B}(\vb*{x}), & \sigma_\mre &= \tilde{\sigma}_\mre \mathcal{B}(\vb*{x}),
\end{align}
where $\mathcal{B}(\vb*{x})$ is the transport efficiency or the inverse MacMullin number, while the tilde distinguishes the microscale (or bulk) value of the parameter from the macroscale (or effective) value of the parameter. In principle, $\mathcal{B}(\vb*{x})$ could be a tensor, to account for the anisotropy of the porous matrix. Here, we treat the efficiency, $\mathcal{B}(\vb*{x})$, as a placeholder and, in the literature, we find several options to model this factor; these range from an entirely systematic approach to computing this quantity from the exact microstructure \cite{Hunt2020,Richardson2012} to \textit{ad hoc} methods based on observed scalings of this factor with $\varepsilon$ (e.g. Bruggeman correlation \cite{Bruggeman1935,Tjaden2016} or Rayleigh method \cite{Bruna2015,Rayleigh1892}). Interestingly, the systematic approach adopted in \cite{Hunt2020,Richardson2012} can be shown to be equivalent to the approach adopted in \cite{LeHoux2021} in which this factor, which they write as $\varepsilon/\tau$ (where $\tau$ is termed the tortuosity factor), is computed from solution of Laplace's equation, with an appropriate boundary conditions, on a domain generated from tomography data of the electrode.

\subsection{Doyle-Fuller-Newman model (DFN)}
\label{sec:DFN}

\begin{figure}
    \centering
    \includegraphics[width=\textwidth]{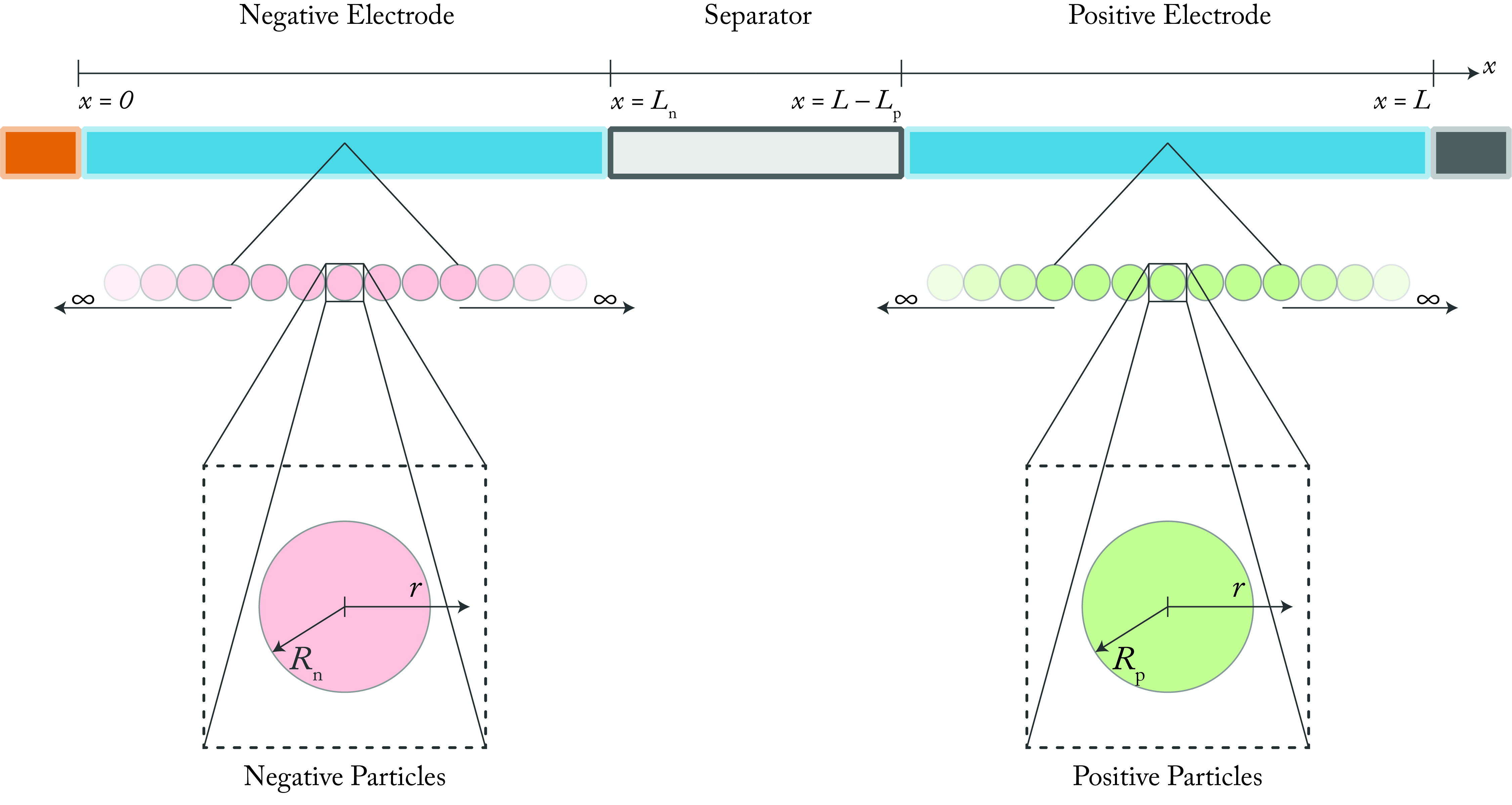}
    \caption{Sketch of the geometry of the DFN model. The electrodes and separator are now one-dimensional domains, defined to be $x \in [0,L_\mrn]$ for the negative electrode, $x \in [L_\mrn, L - L_\mrp]$ for the separator and $x \in [L - L_\mrp,L]$ for the positive electrode. The microstructure is now assumed to be composed of isolated spherical particles, so the domains are $r \in [0, R_k]$ for $\kin{n,p}$.}
    \label{fig:sketch_DFN}
\end{figure}

The Doyle-Fuller-Newman (DFN) model, also known as the pseudo-two-dimensional (P2D) or Newman model, is probably the most popular, physics-based model for lithium-ion batteries. Since the DFN model was first posed in \cite{Fuller1994} this model, and its variants, have been widely used in many different applications \cite{Arora1999,Castle2021,Dees2009,Fuller1994a,Fuller1994,Krachkovskiy2018} and has been the basis of multiple extensions \cite{Srinivasan2003,Yang2017}.

The power of the DFN model, and the main difference from the homogenised model described in Section \ref{sec:homogenisation}, stems from the simple geometry on which it is solved, as shown in Figure \ref{fig:sketch_DFN}. The model assumes that the electrode particles are spherical. The microstructure problem \eqref{eq:homogenised_particle} is thus posed in a one-dimensional domain with spherical symmetry, as opposed to the three-dimensional microstructure from the general homogenised model. In addition, the electrodes and separator are assumed to have a one-dimensional planar geometry, so that \eqref{eq:homogenised_electrode} and \eqref{eq:homogenised_electrolyte} can be reduced to their one-dimensional, planar form. This renders the model simple enough to be computationally affordable, while retaining enough of the physics to be able to accurately predict the batteries' behaviour and capture its internal states. Moreover, given that the microstructure of the electrodes is usually unknown, it provides a simple way of characterising it using an effective microstructure, with relatively few parameters (porosity and particle radius, with the surface area per unit volume calculated from them).

In what follows we state the governing lithium-ion and charge conservation equations, and the related transport equations, that comprise the DFN model. The geometry in which this model is posed is illustrated in Figure \ref{fig:sketch_DFN}. For simplicity, we drop the tildes from the microscopic variables, even though both homogenised and microscopic quantities coexist within the model. We also denote the fluxes as scalars since the microscale lithium-ion flux (in the electrode particles) lies entirely in the radial direction $r$ and the averaged (macroscale) lithium-ion fluxes, and current densities, lie entirely in the $x$-direction.

\paragraph{Particle equation} Conservation of lithium ions in the spherical electrode particles combined with a constitutive equation that represents a (nonlinear) diffusive ion flux, results in the following microscale problems for the lithium-ion concentrations $c_k(r,x,t)$
\begin{subequations}\label{eq:DFN_particle}
\begin{align}\label{eq:DFN_electrode_diffusion}
    \pdv{c_k}{t} +\frac{1}{r^2}\pdv{}{r}\left( r^2 N_k \right)&= 0, & N_k &= -D_k (c_k) \pdv{c_k}{r}, & \mbox{in} \quad & 0 \leq r \leq R_k,
\end{align}
which are solved in the region $0 < x < L_\mrn$ for $k = \mrn$ (negative electrode); and the region $L-L_\mrp < x < L$ for $k = \mrp$ (positive electrode). Here $N_k$ is the radial flux of lithium ions in the active material, $D_k (c_k)$ is the lithium-ion diffusivity in the active material, $r$ is the radial spatial coordinate and $R_k$ is the particle radius. We assume that the particle is entirely surrounded by electrolyte and that lithium transfer within the electrolyte occurs uniformly across each particle's outer surface, which leads to the boundary conditions
\begin{align}
  N_k &= 0, & \text{ on } r = 0,\\
  N_k &= \frac{j_k(x,t)}{F}, & \text{ on } r = R_k,  
\end{align}
where $j_k$ is the interfacial current density. Furthermore, we assume that the concentration within the particles in each electrode is initially uniform in space
\begin{align}
    c_k(r,x,t) &= c_{k0}, & \text{ at } t = 0,
\end{align}
where $c_{k0}$ is a constant. 
\end{subequations}

The interfacial current density, $j_k(x,t)$, (for $\kin{n, p}$) is then given by
\begin{subequations}
\begin{align}
    j_k &= j_{k0} \sinh\left(\frac{F}{2 R T} \eta_k \right), \\
    j_{k0} &= \left. F K_k \sqrt{\frac{c_\mre}{c_{\mre 0}} \frac{c_k}{c_k^{\max}} \left(1 - \frac{c_k}{c_k^{\max}} \right) } \right|_{r=R_k}, \\
    \eta_k &= \phi_k - \phi_e - U_k \left( \left. c_k \right|_{r = R_k} \right),
\end{align}
\end{subequations}
where $c_\mre(x,t)$ is the concentration of lithium ions in the electrolyte, $j_{k 0} (c_k, c_\mre)$ is the exchange current density, $\eta_k(x,t)$ is the surface overpotential and $U_k(c_k)$ is the open-circuit potential.

\paragraph{Electrode equation} Charge conservation and Ohm's Law in the electrode matrices (for $\kin{n, p}$) is described by
\begin{subequations}\label{eq:DFN_phik}
\begin{align}
    \pdv{i_k}{x} &= -b_k j_k,  & i_k &= - \sigma_k \pdv{\phi_k}{x},
\end{align}
where $i_k$ is the averaged macroscopic current density in the electrode matrix, $\phi_k$ is the electric potential in the electrode matrix, $\sigma_k$ is the effective electronic conductivity of the porous electrode matrix, $b_k$ is the surface area per unit volume of electrode particles in contact with the electrolyte. The term $- b_k j_k$ accounts for transfer of current between the electrode and electrolyte, which occurs via the surface intercalation reaction. The domains of the negative and positive electrodes are $0 < x < L_\mrn$ and $L - L_\mrp < x < L$, respectively. Without loss of generality the electric potential is set to zero on the left-hand current collector
\begin{align}
    \phi_\mrn &= 0, & \text{ on } x = 0.
\end{align}
The separator is assumed to be a good insulator so that no current flows directly through the separator matrix, from one electrode to the other, and thus
\begin{align}
    i_\mrn &= 0, & \text{ on } x = L_\mrn,\\
    i_\mrp &= 0, & \text{ on } x = L - L_\mrp.
\end{align}
For galvanostatic charge/discharge the remaining boundary condition is provided by specifying the current flow at one of the current collectors
\begin{align}\label{eq:current_bc}
i_\mrp &= i_\mathrm{app}(t), & \text{ on } x = L,
\end{align}
where $i_\mathrm{app}(t)$ is the applied current density. Alternatively, for potentiostatic charge/discharge, the potential difference  between the current collectors $V(t)$ is specified and, since we have already specified $\phi_\mrn|_{x=0}=0$, it corresponds to the condition
\begin{align}\label{eq:DFN_prescribed_V_BC}
\phi_\mrp &= V(t), & \text{ on } x = L.
\end{align}
\end{subequations}

More complicated modes, such as power or resistance control, can be modelled using the boundary condition \eqref{eq:current_bc}, together with an additional algebraic equation that relates the current flow in the cell  ($I = A i_\mathrm{app}$, where $A$ is the electrode plate area) to $V(t)$ the potential difference across the cell. For example,
\begin{equation}
    I(t) V(t) = P(t)
\end{equation}
describes a cell with power draw $P(t)$.

\paragraph{Electrolyte equations}  The electrolyte occupies the entire region between the current collectors, $0 \leq x \leq L$, and in this region the averaged current density, $i_\mre$, satisfies a current conservation equation and a constitutive equation analogous to Ohm's Law but for an electrolyte. These read
\begin{subequations}\label{eq:DFN_phie}
\begin{align}
    \pdv{i_\mre}{x} &=  b(x) j(x), &
    i_\mre = \sigma_\mre (c_\mre) \mathcal{B}(x) \left( -  \pdv{\phi_\mre}{x} + \frac{2}{F} (1-t^+) \dv{\mu_\mre}{c_\mre} \pdv{c_\mre}{x} \right),
\end{align}
in which
\begin{align}
    b (x) &= \begin{cases}
    b_\mrn, & \text{ if } 0 \leq x \leq L_\mrn,\\ 
    b_\mrs, & \text{ if } L_\mrn < x \leq L - L_\mrp,\\ 
    b_\mrp, & \text{ if } L - L_\mrp < x \leq L,
    \end{cases} & 
    j (x) &= \begin{cases}
    j_\mrn, & \text{ if } 0 \leq x \leq L_\mrn,\\ 
    0, & \text{ if } L_\mrn < x \leq L - L_\mrp,\\ 
    j_\mrp, & \text{ if } L - L_\mrp < x \leq L,
    \end{cases} &
    \mathcal{B} (x) &= \begin{cases}
    \mathcal{B}_\mrn, & \text{ if } 0 \leq x \leq L_\mrn,\\ 
    \mathcal{B}_\mrs, & \text{ if } L_\mrn < x \leq L - L_\mrp,\\ 
    \mathcal{B}_\mrp, & \text{ if } L - L_\mrp < x \leq L.
    \end{cases}
\end{align}
Here $b(x)$ is the surface area per unit volume of the electrode, $j(x)$ is the interfacial current density caused by the intercalation reaction, $\sigma_\mre$ is the electrolyte conductivity, $\mathcal{B}(x)$ is the transport efficiency, $\phi_\mre$ is the electrical potential in the electrolyte and $t^+$ is the transference number. In the electrodes, charge is transferred between the electrolyte and the electrode matrix, whereas in the separator, no charge transfer occurs between the electrolyte and the separator material. Since no charge is transferred directly from the electrolyte to the current collectors the following conditions are satisfied on the edges of the domain:
\begin{align}
    i_\mre &= 0, & \text{ at } x = 0, L.
\end{align}
\end{subequations}

Conservation of lithium-ions in the electrolyte and an appropriate constitutive equation for the averaged lithium ion flux $N_\mre$ (as derived in \cite{Newman2004,Richardson2021}), give the following problem for the lithium-ion concentration $c_\mre$ in $0 \leq x \leq L$:
\begin{subequations}
\begin{align}
  \varepsilon(x) \pdv{c_\mre}{t} &= -\pdv{N_\mre}{x} + \frac{b(x) j(x)}{F}, & N_\mre &= -D_\mre (c_\mre) \mathcal{B}(x) \pdv{c_\mre}{x} + \frac{t^+}{F} i_\mre,
\end{align}
with
\begin{align}
    \varepsilon (x) &= \begin{cases}
    \varepsilon_\mrn, & \text{ if } 0 \leq x \leq L_\mrn,\\ 
    \varepsilon_\mrs, & \text{ if } L_\mrn < x \leq L - L_\mrp,\\ 
    \varepsilon_\mrp, & \text{ if } L - L_\mrp < x \leq L.
    \end{cases}
\end{align}
where $\varepsilon_{k}$ is the electrolyte volume fraction and $D_\mre (c_\mre)$ is the bulk diffusivity of the electrolyte. Since no flux of lithium ions flows from the electrolyte into the current collectors the boundary conditions on the edges of the domain are
\begin{align}
    N_\mre &= 0, & \text{ at } x = 0,L.
\end{align}
Finally, we assume that the lithium-ion concentration in the electrolyte is initially uniform in space, so that
\begin{align}
    c_\mre &= c_{\mre 0}, & \text{ at } t = 0,
\end{align}
\end{subequations}
where $c_{\mre 0}$ is constant. 

\subsubsection{Battery energy balance for the DFN model \label{sec:econs}}
Lithium-ion batteries are energy storage devices and thus an important measure of their performance is the fraction of the stored chemical energy that can be recovered from the device as useful electrical energy. The fraction of the chemical energy that is not converted into useful electrical work as the device is discharged is converted into heat and is known as irreversible energy loss. This generated heat, as discussed in Section \ref{sec:thermal}, is key for the battery's behaviour, which is highly sensitive to changes in temperature. However, for now, we restrict the scope of the discussion to the energy balance. The total energy balance for the DFN model takes the form
\begin{equation}\label{econs}
-A \dv{G}{t} = I V + A \dot Q_\mathrm{irr}, 
\end{equation}
where $G$ is the Gibbs free energy per unit area of the cell, $A$ is the cell's area, $\dot Q_\mathrm{irr}$ is the energy dissipated per unit area and $I$ and $V$ are the current drawn by and the potential drop across the cell. The physical interpretation of this law is that the left-hand side of \eqref{econs} represents the rate of change of the total chemical energy within the cell, while the terms on the right-hand side represent the rate of energy drawn from the cell. In particular, the terms on the right-hand side can be sub-divided into $I V$, the useful electrical power produced by the device, and $A \dot Q_\mathrm{irr}$, the rate of irreversible energy dissipation (to heat). Further details, on the computation of $G$ and $\dot Q_\mathrm{irr}$ from the solution of the DFN model, together with a mathematical proof of this result and its numerical validation, can be found in \cite{Richardson2021heat}. This formulation provides all the contributions to the irreversible heat generation, which need to be considered in coupled thermal-electrochemical models of a battery. A similar consistent theory has been proposed for the microscopic model by Latz \& Zausch \cite{Latz2011,Latz2015}.

\section{Doyle-Fuller-Newman model to Single Particle Model}\label{sec:DFN_to_SPM}

Even though the DFN model is itself a simplified, homogenised model of the complex three-dimensional microscopic effects occurring in batteries, it is still quite complex and requires carefully designed numerical algorithms, in order to efficiently obtain accurate numerical solutions. The computational costs of solving the DFN model becomes particularly important when the model is being used as a tool to optimise battery design, estimate electrochemical parameters of a battery, or implement accurate real-time controllers on the relatively modest computational resources found in electric vehicles. However, perhaps the most computationally intensive use that these models can be applied to is as part of a thermally coupled, electrochemical model of large format batteries (e.g. pouch and cylindrical batteries). Such models give rise to problems which, in addition to the two spatial dimensions found in the DFN model, are also coupled to an additional three macroscopic spatial dimensions, accounting for the heat flow across the large format battery. The resulting five-dimensional problem is extremely computationally complex and requires either powerful computational resources or highly advanced numerical methods (e.g. \cite{Korotkin2021}) to solve to a satisfactory degree of accuracy.

Reduced order models, i.e.  models that are simpler than the DFN model but retain most of its predictive capabilities, can improve physical understanding and provide accurate solutions at a significantly lower computational cost. The most popular family of reduced order models are the Single Particle Models (SPMs). The key idea of SPMs is that the behaviour of the particles within each electrode is very similar and, therefore, they can all be approximated by a single representative (or average) particle, hence the name. The foundations of the Single Particle Model were first introduced in 1979 by Atlung et al. \cite{Atlung1979} considering only the mass transport in three different particle geometries: planar, cylindrical and spherical. Recently, these models have garnered renewed interest from the battery modelling community, with several authors presenting different versions of SPMs, with and without electrolyte dynamics. There are two main approaches to formulating these models. While some authors take an \textit{ad hoc} approach, posing the model directly from a list of simplifying assumptions \cite{Bizeray2018,Han2015i,Han2015ii,kemper2013extended,Moura2017a}, others (particularly in the mathematical modelling community) use formal asymptotic methods to systematically derive the reduced order models directly from the DFN model. Over the past few years, several authors have applied asymptotic methods to the DFN model to derive reduced order battery models \cite{BrosaPlanella2021,Marquis2019,Moyles2019,Richardson2012,Richardson2020}. The exact form of the model that results from this process depends upon the underlying assumptions on which the asymptotic analysis is based.

These different approaches have led to a number of similar, but distinct, reduced order models, all of which are termed Single Particle Models. This can make navigating the literature a daunting task, particularly since they all share a similar structure and some fundamental characteristics, independently of how they have been derived. It is important to clarify the nomenclature at this point. We group all these models under the definition of Single Particle Model type (SPM-type), which refers to the entire family of models that use a single particle to represent all the particles within a given electrode. Within this family, we distinguish two subfamilies: models which include electrolyte dynamics (which we refer to as Single Particle Models with electrolyte, or SPMe) and models which do not (which we refer to as Single Particle Models, or SPM). When we refer to a particular model from a specific reference, we use the article details, regardless of how the authors named the model, to keep consistency across this review article (e.g. Marquis et al. \cite{Marquis2019} or Richardson et al. \cite{Richardson2020}, which are two different instances of SPMe).

Crucially, for SPM-type models the PDEs in the microscopic spatial variable, $r$, decouple from those in the macroscopic spatial variable, $x$. This results in a problem that is effectively one-dimensional in space, as opposed to the DFN model, which is genuinely two-dimensional in space. It is this aspect of these models that is mainly responsible for their much reduced computational complexity when compared to the DFN model. The main idea behind this decoupling is that, in many circumstances, the intercalation reaction occurs almost uniformly across both electrodes. That is to say that during discharge, all negative electrode particles delithiate at (almost) the same rate, independently of their position in the negative electrode, and all positive electrode particles lithiate at (almost) the same rate, independently of their position in the positive electrode (and similarly for battery charge). It follows that, in such scenarios, their behaviour is well approximated by a single representative particle in each electrode. This holds for most materials, however lithium iron phosphate is a notable exception due to its flat open-circuit potential, and requires a different approach (see \cite{Castle2021}). In scenarios where the current applied to the battery is given as part of the problem formulation, the current density on the surfaces of the electrode particles can be computed in advance. This leads to a decoupling between the PDEs for potential and lithium concentration in the electrodes and electrolyte. In many cases, the equations for the potential in the electrodes and the electrolyte can be solved analytically in advance, leaving only the equations for the lithium concentration in the electrode particles and the electrolyte to be solved numerically. In more complex scenarios, where these analytical solutions are not possible, the potential equations can be solved numerically but at a much lower computational cost than solving the fully coupled DFN model.

The spatial uniformity in discharge can either be assumed directly \cite{Moura2017a} or derived systematically, using asymptotic methods, based either on very slow battery discharge \cite{Marquis2019} or on rapid changes in the open-circuit potentials (measured in terms of the thermal potential), with respect to changes in the lithium concentrations within the electrode materials \cite{Ranom2014,Richardson2020}. The main advantages of these methods is that they offer a systematic approach to model reduction and are therefore applicable to modified versions of the DFN model, and that they ensure consistency of the reduced order model with the underlying model, allowing for the assumptions to be validated and the errors estimated \textit{a priori}. There are further simplifications that can be applied to these models, such as fast lithium transport in the electrode particles, see e.g. \cite{Moyles2019,Richardson2012}, but we do not discuss these here, and instead refer the reader to \cite{BrosaPlanella2021}.

\subsection{Single Particle Models with electrolyte dynamics (SPMe)}\label{sec:SPMe-type}
Several models in the literature fit under the category of Single Particle Models with electrolyte dynamics, despite receiving multiple names. But a closer inspection shows that the main differences between these  models are in the methods used to calculate the voltage and other derived quantities, not in the differential equations that we need to solve, in order to determine the concentrations in the particles and in the electrolyte.

\begin{figure}
    \centering
    \includegraphics[width=\textwidth]{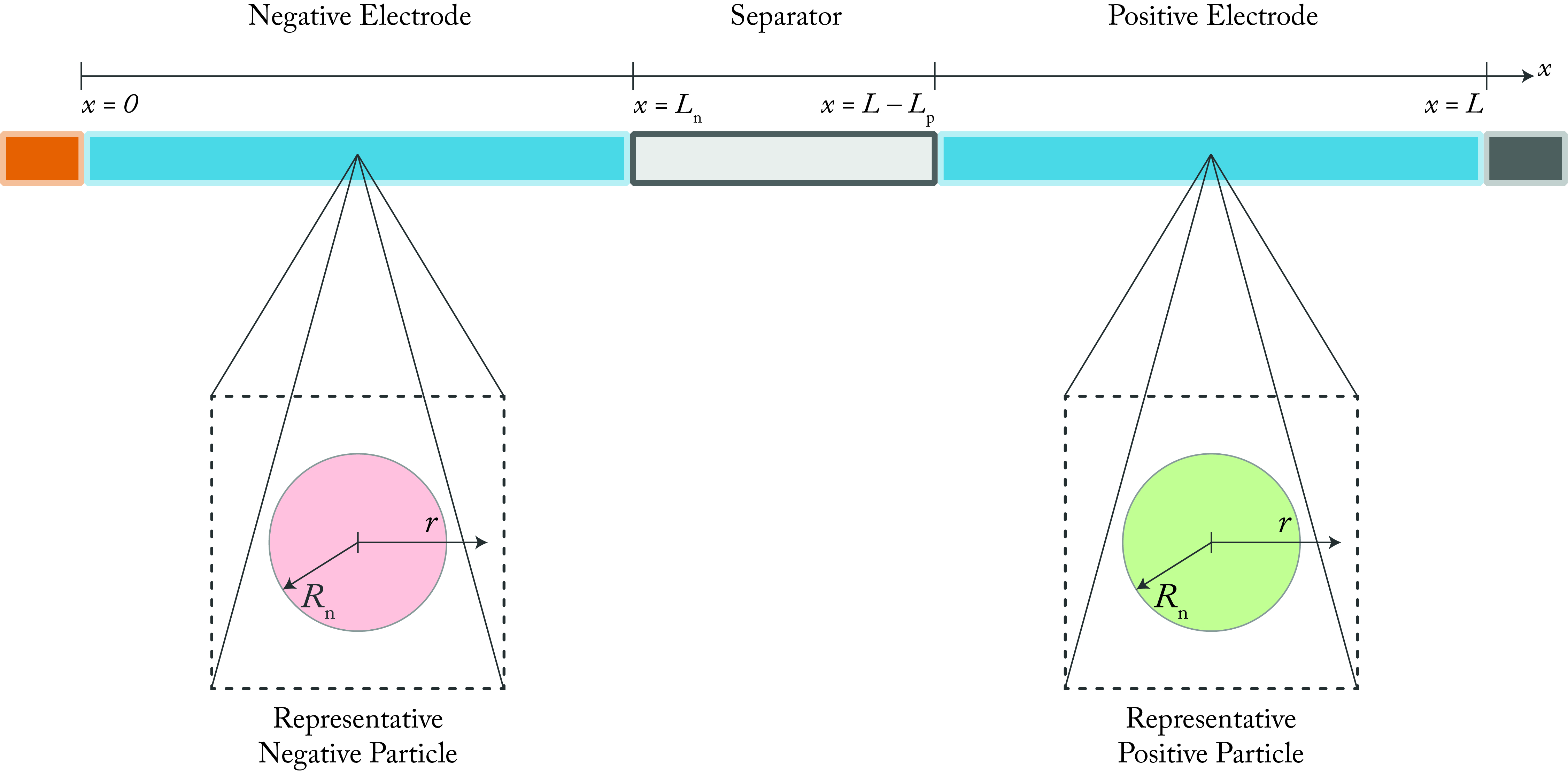}
    \caption{Sketch of the geometry of the SPMe model. The geometry is very similar to that of the DFN model, but now there is only a single representative particle at each electrode, rather than infinitely many.}
    \label{fig:sketch_SPMe}
\end{figure}

As explained earlier, in SPM-type models the particles are assumed to (de)lithiate uniformly and therefore we can solve for a representative particle in each electrode. The governing equations for these representative particles are
\begin{subequations}\label{eq:SPMe2_particle}
\begin{align}
\pdv{c_{k}}{t} &= \frac{1}{r^2} \pdv{}{r} \left( r^2 D_{k}(c_{k}) \pdv{c_{k}}{r} \right), 
&& \text{ in } 0 < r < R_k,\\
\pdv{c_{k}}{r} &= 0, && \text{ at } r = 0,
\\
-D_{k}(c_{k}) \pdv{c_{k}}{r} &= \frac{j_k}{b_k F}, && \text{ at } r = R_k,\\
c_{k} &= c_{k0}, && \text{ at } t = 0,
\end{align}
for $\kin{n, p}$, and where
\begin{align}\label{eq:SPMe2}
j_\mrn &= \frac{i_\mathrm{app}(t)}{L_\mrn}, & j_\mrp &= -  \frac{i_\mathrm{app}(t)}{L_\mrp}.
\end{align}
\end{subequations}

The SPMe also accounts for electrolyte effects, so we need to solve the governing equation for lithium ions in the electrolyte, which reads
\begin{subequations}\label{eq:SPMe_electrolyte}
\begin{align}
\varepsilon (x)\pdv{c_\mre}{t} &= \pdv{}{x} \left(D_{\mre}(c_\mre) \mathcal{B}(x) \pdv{c_\mre}{x} - \frac{t^+}{F} i_\mre \right) + \frac{b(x) j(x,t)}{F}, && \text{ in } 0 < x < L,\\
\pdv{c_\mre}{x} &= 0, && \text{ at } x = 0, L,\\
c_{\mre} &= c_{\mre 0}, && \text{ at } t = 0.
\end{align}
where
\begin{equation}
\begin{aligned}
    \varepsilon(x) &= \begin{cases}
    \varepsilon_\mrn, & \text{ in } 0 \leq x \leq L_\mrn, \\
    \varepsilon_\mrs, & \text{ in } L_\mrn \leq x \leq L - L_\mrp, \\
    \varepsilon_\mrp, & \text{ in } L - L_\mrp \leq x \leq L,
    \end{cases} & 
    \mathcal{B}(x) &= \begin{cases}
    \mathcal{B}_\mrn, & \text{ in } 0 \leq x \leq L_\mrn, \\
    \mathcal{B}_\mrs, & \text{ in } L_\mrn \leq x \leq L - L_\mrp, \\
    \mathcal{B}_\mrp, & \text{ in } L - L_\mrp \leq x \leq L,
    \end{cases} \\
    b(x) &= \begin{cases}
    b_\mrn, & \text{ in } 0 \leq x \leq L_\mrn, \\
    b_\mrs, & \text{ in } L_\mrn \leq x \leq L - L_\mrp, \\
    b_\mrp, & \text{ in } L - L_\mrp \leq x \leq L.
    \end{cases} &
    j(x,t) &= \begin{cases}
    j_\mrn(t), & \text{ in } 0 \leq x \leq L_\mrn, \\
    0, & \text{ in } L_\mrn \leq x \leq L - L_\mrp, \\
    j_\mrp(t), & \text{ in } L - L_\mrp \leq x \leq L,
    \end{cases}\\
    i_\mre(x,t) &= \begin{cases}
    \frac{i_\mathrm{app}(t) }{L_\mrn} x, & \text{ in } 0 \leq x \leq L_\mrn, \\
    i_\mathrm{app}(t), & \text{ in } L_\mrn \leq x \leq L - L_\mrp, \\
    \frac{i_\mathrm{app}(t)}{L_\mrp} (L - x), & \text{ in } L - L_\mrp \leq x \leq L,
    \end{cases}
\end{aligned}
\end{equation}
\end{subequations}

The solution for $c_\mre(x,t)$ can then be used to compute the electrolyte potential $\phi_e$ \textit{a posteriori}, by integrating the expression
\begin{align}\label{eq:SPMe_electrolyte_potential}
    i_\mre(x,t) = \sigma_\mre (c_\mre) \mathcal{B}(x) \left( -  \pdv{\phi_\mre}{x} + \frac{2 R T}{F} (1-t^+)  \frac{1}{c_\mre} \pdv{c_\mre}{x} \right) \qquad \text{ in } 0 < x < L.
\end{align}
Here we have assumed the electrolyte is ideal such that $\mu_\mre = RT \log (c_\mre/c_{\mre 0})$.\footnote{In ~\cite{Richardson2020} this integration is accomplished numerically, in~\cite{BrosaPlanella2021} it is performed analytically, and in~\cite{Marquis2019} an approximation based on a near constant concentration $c_\mre \approx c_{\mre0}$ is used.}

Multiple methods have been suggested, in the literature, to compute the terminal voltage of the cell. Here we focus on the works of Marquis et al.~\cite{Marquis2019} and Richardson et al.~\cite{Richardson2020}, which both show very good agreement to solutions of the DFN model. We describe their results in what follows. In both cases the terminal voltage $V(t)$ can be written in the form
\begin{equation}\label{eq:V_SPMe_general}
    V(t) = U_\mathrm{eq} - \eta_\mathrm{r} - \eta_\mathrm{c} - \Delta \phi_\mre - \Delta \phi_\mrs,
\end{equation}
where $U_\mathrm{eq}$ is the open-circuit potential of the cell, $\eta_\mathrm{r}$ and $\eta_\mathrm{c}$ are the potential drops due to the reaction and concentration overpotentials, respectively; and $\Delta \phi_\mre$ and $\Delta \phi_\mrs$ are the Ohmic losses in the electrolyte and the electrodes, respectively. During discharge ($i_\mathrm{app} > 0$), as lithium ions flow from the negative to positive electrode, the quantities $\eta_\mathrm{r}$, $\eta_\mathrm{c}$,  $\Delta \phi_\mre$ and $\Delta \phi_\mrs$ are all positive, and represent the reduction in the open circuit cell potential resulting from these loss mechanisms. During charge ($i_\mathrm{app}<0$) these quantities are all negative and therefore represent the extra potential, above the open-circuit potential, required to overcome the internal resistances in the cell.  This decomposition of the voltage into constituent parts naturally appears when the model is derived (see, for example, \cite{Marquis2019}), thus providing further physical intuition into what effects govern the voltage response of the battery.

\begin{table}[htbp]
    \centering
    \begin{tabular}{| p{0.6cm} | p{6.9cm} | p{6.7cm} |}
        \cline{2-3} 
        \multicolumn{1}{c |}{} & \multicolumn{1}{c |}{\textbf{Marquis et al. \cite{Marquis2019}}} & \multicolumn{1}{c |}{\textbf{Richardson et al. \cite{Richardson2020}}} \\ 
        \hline
        $$U_\mathrm{eq}$$ & \multicolumn{2}{ p{13cm} |}{$$U_\mrp \left( \left. c_{\mrp} \right|_{r = R_\mrp} \right) - U_\mrn \left( \left. c_{\mrn} \right|_{r = R_\mrn} \right)$$}\\ 
        \hline
        $$\eta_\mathrm{r}$$ & $$\frac{2 R T}{F} \left(\arcsinh \left( \frac{j_\mrn(t)}{\bar{j}_{\mrn 0}(t)} \right)-\arcsinh \left( \frac{j_\mrp(t)}{\bar{j}_{\mrp 0}(t)} \right) \right)~$$ & 
        $$\begin{aligned}  \frac{2 R T}{F} \left( \frac{1}{L _\mrn} \int_{0}^{L_\mrn} \arcsinh \left( \frac{j_\mrn(t)}{j_{\mrn 0}(x,t)} \right) \dd x \right. \\ 
        \left. - \frac{1}{L _\mrp} \int_{L - L_\mrp}^{L} \arcsinh \left( \frac{j_\mrp (t)}{j_{\mrp 0}(x,t)} \right) \dd x \right) \end{aligned}$$ \\ 
        \hline
        $$\eta_\mathrm{c}$$ & $$\begin{aligned} (1 - t^+) \frac{2 R T}{F c_{\mre 0}} \left(  \frac{1}{L _\mrn} \int_{0}^{L_\mrn} c_\mre (x,t) \dd x \right.  \\  \left. -\frac{1}{L _\mrp} \int_{L - L_\mrp}^{L} c_\mre (x,t) \dd x  \right)\end{aligned}$$ & 
        $$\begin{aligned}
        (1-t^+) \frac{2 R T}{F} \left( \frac{1}{L _\mrn} \int_{0}^{L_\mrn} \log \left( c_\mre (x,t) \right) \dd x  \right. \\
        \left. -  \frac{1}{L _\mrp} \int_{L - L_\mrp}^{L} \log \left( c_\mre (x,t) \right) \dd x \right)
        \end{aligned}$$
        \\ 
        \hline
        $$\Delta \phi_\mre$$ & $$\frac{i_\mathrm{app}}{\sigma_\mre(c_{\mre 0})} \left(\frac{L _\mrn}{3 \mathcal{B}_\mrn} + \frac{L_\mrs}{\mathcal{B}_\mrs} + \frac{L _\mrp}{3 \mathcal{B}_\mrp}\right)$$ & 
        $$\begin{aligned} \left(\frac{1}{L _\mrp} \int_{L - L_\mrp}^{L} \int_0^x \frac{i_\mre(s,t) \dd s}{\sigma_\mre \left(c_\mre(s,t)\right) \mathcal{B}(s)} \dd x \right. \\
        \left. - \frac{1}{L _\mrn} \int_0^{L_\mrn} \int_0^x \frac{i_\mre(s,t) \dd s}{\sigma_\mre \left(c_\mre(s,t)\right) \mathcal{B}(s)} \dd x \right)
        \end{aligned}$$\\ 
        \hline
        $$\Delta \phi_\mrs$$ & \multicolumn{2}{p{13cm} |}{$$ \frac{i_\mathrm{app}}{3} \left( \frac{L _\mrp}{\sigma_\mrp} + \frac{L _\mrn}{\sigma_\mrn}\right)$$} \\
        \hline
    \end{tabular}
    \caption{Contributions to the terminal voltage \eqref{eq:V_SPMe_general} for the models presented in Marquis et al. \cite{Marquis2019} and Richardson et al. \cite{Richardson2020}. The exact expressions are different from those used in the original papers to keep them consistent with the models and notation in this review.}
    \label{tab:SPMe_voltage_components}
\end{table}

The expressions for each contribution according to each paper are shown in Table \ref{tab:SPMe_voltage_components}. The exchange current densities are defined as
\begin{align}
j_{k0}(x,t) &= F K_k \left. \sqrt{\frac{c_{\mre}(x,t)}{c_{\mre 0}} \frac{c_{k}(r,t)}{c_k^{\max}} \left(1 - \frac{c_{k}(r,t)}{c_{k}^{\max}} \right)} \right|_{r = R_k} \qquad \mbox{for} \quad \kin{n,p}.
\end{align} 
In turn, $\bar{j}_{k0}(t)$ is defined as the exchange current density averaged over the corresponding electrode.

Note that the model in Richardson et al.~\cite{Richardson2020} (which the authors term the corrected Single Particle Model, or cSPM) is derived using asymptotic methods based on the disparity between the size of the thermal potential and the characteristic change in overpotential occurs as the electrode materials are (de)lithiated. It also allows for graded electrodes, in which the porosity, particle radius, or surface area change with position across the electrodes. However, here we restrict our attention to the uniform electrodes problem only (and provide the corresponding expressions), while we briefly discuss these extensions to the model in Section \ref{sec:extensions}. In the uniform electrodes problem we can use the explicit expressions shown in Table \ref{tab:SPMe_voltage_components}, and the details on the derivation follow very similarly to the analysis in \cite{BrosaPlanella2021}. In the graded electrode case, we need to solve \eqref{eq:SPMe_electrolyte_potential} numerically. Then, the two electrolyte contributions $\eta_\mathrm{c}$ and $\Delta \phi_\mre$ are grouped together in the same contribution, given by
\begin{equation}
    \eta_\mathrm{c} + \Delta \phi_\mre =  \frac{1}{L _\mrn} \int_0^{L_\mrn} \phi_\mre (x, t) \dd x-\frac{1}{L _\mrp} \int_{L - L_\mrp}^{L} \phi_\mre (x, t) \dd x.
\end{equation}
The model presented in Marquis et al.~\cite{Marquis2019} (which the authors term the canonical SPMe), has been derived using asymptotic methods taking the assumptions of fast ion transport in the electrolyte and high conductivity both in the electrodes and the electrolyte.

Comparing the contributions to the terminal voltage shown in Table \ref{tab:SPMe_voltage_components} shows considerable similarity between the two approaches. The key difference between both models is in the electrolyte concentration. The model in Marquis et al. finds that the deviations from the electrolyte initial concentration are small, while the model in Richardson et al. allows for significant variations in the electrolyte concentration at leading order. It is this difference that leads to differences in the terms $\eta_\mathrm{r}$, $\eta_\mathrm{c}$ and $\Delta \phi_\mre$. As shown in \cite{Timms2020Corrigendum}, for the particular example of the battery parameterised by Ecker et al. \cite{Ecker2015i,Ecker2015ii}, the performance of the two models is very similar with differences only appearing at high C-rates, where the accuracy of the model in Richardson et al. \cite{Richardson2020} is slightly better.

\subsection{Single Particle Model (SPM)}
\label{sec:SPM}

The Single Particle Model (SPM) is a further simplification of the SPMe in which the potential drops across the electrolyte are neglected. In the limit of very low C-rate, gradients in both the concentration and the potential of the electrolyte are negligible, and the SPM model can be applied with relative accuracy. The SPM consists of only two diffusion equations, one for the representative particle of each electrode, and a closed-form expression for the voltage.

The governing equation for the electrode particles is \eqref{eq:SPMe2_particle}. Then, the terminal voltage can be written in the form

\begin{subequations}\label{eq:SPM_voltage}
\begin{equation}
    V = U_\mathrm{eq} - \eta_\mathrm{r},
\end{equation}
where
\begin{align}
    U_\mathrm{eq} &= U_\mrp \left( \left. c_{\mrp} \right|_{r = R_\mrp} \right) - U_\mrn \left( \left. c_{\mrn} \right|_{r = R_\mrn} \right),
    && (\text{open-circuit potential}),
    \\
    \eta_\mathrm{r} &= \frac{2 R T}{F} \arcsinh \left( \frac{j_\mrn(t)}{j_{\mrn 0}(t)} \right)-\frac{2 R T}{F} \arcsinh \left( \frac{j_\mrp(t)}{j_{\mrp 0}(t)} \right),
    && (\text{reaction overpotential}),
\end{align}
\end{subequations}
with
\begin{align}
j_{k0}(t) &= F K_k \left. \sqrt{\frac{c_{k}(r,t)}{c_k^{\max}} \left(1 - \frac{c_{k}(r,t)}{c_{k}^{\max}} \right)} \right|_{r = R_k} \qquad \mbox{for} \quad \kin{n,p}.
\end{align}
This expression is a simplification of \eqref{eq:V_SPMe_general}, obtained by taking $c_\mre = c_{\mre 0}$  in the SPMe models of Marquis et al. \cite{Marquis2019} and Richardson et al. \cite{Richardson2020}. This represents a considerable reduction in complexity because it eliminates  the need to solve an additional differential equation for the electrolyte concentration.

Note also that in the case where the Butler-Volmer equation is not symmetric, we cannot directly invert the Butler-Volmer equation to obtain a closed form for $\eta_\mathrm{r}$ given the interfacial current density. Instead, we need to determine $\eta_\mathrm{r}$ by solving a nonlinear algebraic equation (see details in \cite{Richardson2012}). However, the structure of the model remains the same (for both SPM and SPMe).



\section{Selecting an appropriate model}\label{sec:comparison}

In the previous two sections we have demonstrated how a set of models, with different levels of complexity, can be systematically derived from the microscale model by using the method of formal asymptotic expansions. This reductive approach eventually leads to the Single Particle Model, perhaps the simplest physics-based model describing the behaviour of a battery. In order to choose an appropriate model for a particular application it is imperative to understand its advantages and limitations. One common constraint is on the computational power required to solve the model equations, but even in scenarios where sufficient computational resource is available it is not always clear that the more complex model is necessarily better. In particular, although more complex models offer potentially greater accuracy, they can cloud the physical insight that can be gained from a simpler model and, as shown in Table \ref{tab:parameters}, they require a more detailed parameterisation which can be a needless overhead if high accuracy is not required \cite{Wang2022}.

This section explains what it is possible to learn about the mechanisms limiting the performance of a lithium-ion battery using each of the models described in the previous sections. Starting from the simplest model, the SPM, we discuss the meaning of the various parameters in the context of simulated behaviours of the cell, as well as what can be gained by adding more complex physics to the description. Crucially, the risks of this additional complexity are highlighted, including the challenges associated with validation and identifiablity.

Before starting with the simple SPM model, we briefly comment on the \textit{so-called} equivalent-circuit models (ECM) \cite{Hu2012,Plett2015,Widanage2016}. These are an even simpler class of model,  widely used in many applications. ECMs assume that the battery can be represented by an electrical circuit, typically comprised of resistors and capacitors, and then fit the parameters of the circuit components to experimental data. They have the advantage of being very cheap to simulate (a small system of ODEs) and quick to parameterise. However, since they are empirical models rather than physics-based models, they offer no insight into the physical processes occurring in the battery. But their simplicity, and the ease with which they are solved,  has led to them being widely used in battery management systems \cite{Farmann2016}. They thus provide a useful reference point to compare other models against, particularly with regards to complexity and the insight that they provide.

The complexity associated with solving a model depends, to a degree, on the solution method employed. A common approach applied to dissipative partial differential equations (PDEs), such as the DFN, SPMe and SPM models, is the \textit{so-called} method of lines \cite{Schiesser2016}. This approach requires only the spatial derivatives to be discretised, leading to a large set of coupled time-dependent ordinary differential equations (ODEs) and algebraic equations that can then be solved using a standard package, such as MATLAB's \texttt{ode15s}. Different techniques can be employed to discretise the spatial operators in the governing PDEs, examples being: finite volume methods \cite{Leveque2002}, control volume methods \cite{Zeng2013}, and finite element methods \cite{Johnson2009,Korotkin2021}. Depending on the nature of the PDEs that comprise the model, different systems of temporal equations arise from the spatial discretisation. If all the PDEs have a time derivative, then we obtain a system of ordinary differential equations (ODEs), while if none has time derivatives we obtain a system of algebraic equations. A mixed scenario is also often encountered in which some equations have time derivatives and others do not; after discretisation this yields a system of differential-algebraic equations (DAEs). Numerical solution of DAEs is more complex than either the solution of a system of ODEs, or that of a system of algebraic equations, and this plays a role in the complexity of battery models that we discuss.


\begin{table}[htbp]
    \centering
    \begin{tabular}{| c | c | c | c | c | c |}
        \hline
        & SPM & SPMe & DFN & Homogenised & Microscale \\ \hline
        $T$ & $\checkmark$ & $\checkmark$ & $\checkmark$ & $\checkmark$ & $\checkmark$\\ \hline
        $R_k$ & $\checkmark$ & $\checkmark$ & $\checkmark$ & * & * \\
        $L_k$ & $\checkmark$ & $\checkmark$ & $\checkmark$ & * & * \\
        $b_k$ & $\checkmark$ & $\checkmark$ & $\checkmark$ & $\checkmark$ & * \\
        $D_k$ & $\checkmark$ & $\checkmark$ & $\checkmark$ & $\checkmark$ & $\checkmark$ \\
        $c_{k0}$ & $\checkmark$ & $\checkmark$ & $\checkmark$ & $\checkmark$ & $\checkmark$ \\
        $K_k$ & $\checkmark$ & $\checkmark$ & $\checkmark$ & $\checkmark$ & $\checkmark$ \\
        $U_k$ & $\checkmark$ & $\checkmark$ & $\checkmark$ & $\checkmark$ & $\checkmark$ \\
        $c_k^{\max}$ & $\checkmark$ & $\checkmark$ & $\checkmark$ & $\checkmark$ & $\checkmark$ \\ 
        $\sigma_k$ & & $\checkmark$ &  $\checkmark$ & $\checkmark$ & $\dagger$ \\ \hline
        $L_\mrs$ & & $\checkmark$ & $\checkmark$ & $\checkmark$ & * \\ \hline
        $\varepsilon$ & & $\checkmark$ &  $\checkmark$ & $\checkmark$ & * \\
        $\mathcal{B}$ & & $\checkmark$ &  $\checkmark$ & $\checkmark$ & * \\
        $D_\mre$ & & $\checkmark$ & $\checkmark$ & $\checkmark$ & $\checkmark$ \\
        $\sigma_\mre$ & & $\checkmark$ & $\checkmark$ & $\checkmark$ & $\checkmark$ \\
        $t^+$ & & $\checkmark$ &  $\checkmark$ & $\checkmark$ & $\checkmark$ \\
        $c_{\mre 0}$ & &$\checkmark$ &  $\checkmark$ & $\checkmark$ & $\checkmark$ \\ \hline
    \end{tabular}
    \caption{Parameters required for each model. $\checkmark$ means that the parameter is needed for that model. * means that the parameter ``as is'' is not needed, but a detailed geometry needs to be provided instead (e.g. when using tomography images for the particles we do not need to provide the radius). $\dagger$ means that the parameter needs to be resolved at the microscale, including material heterogeneities. Note that for the variables with subscript $\kin{n, p}$ we need a different value for each electrode. Many of these parameters, such as $U_k$, $D_k$, $D_\mre$ and $\sigma_\mre$, are usually taken to be functions of the lithium concentration.}
    \label{tab:parameters}
\end{table}

\subsection{Single Particle Models (SPM)}

The Single Particle Model (SPM), as described in Section \ref{sec:SPM}, is the simplest of the models presented in Sections \ref{sec:full_to_DFN} and \ref{sec:DFN_to_SPM}. This model, governed by \eqref{eq:SPMe2_particle} and \eqref{eq:SPM_voltage}, incorporates three basic physical phenomena: lithium transport in the particles \eqref{eq:SPMe2_particle}, a thermodynamic relation between lithium concentration and electrode potential (\ref{eq:SPM_voltage}b), and the overpotential required to drive the lithium intercalation reactions (\ref{eq:SPM_voltage}c). The terminal voltage, as predicted by the SPM, does not include any contributions from the electrolyte (both Ohmic losses and concentration overpotentials), nor any contribution due to Ohmic losses in the electrodes. These contributions are typically only negligible when the cell is operating in a low current regime. 

When implemented numerically, the complexity of the SPM is similar to that of equivalent-circuit models. The low complexity of the SPM is also apparent when considering the number of parameters needed to characterise the model. Apart from the physical constants, $R$ and $F$, and the model input, $i_\mathrm{app}$, we only need 17 parameters to fully characterise the SPM (Table \ref{tab:parameters}): 8 parameters for each electrode, plus the battery temperature. Note that some of these parameters, such as the open-circuit potentials and (usually) the diffusion coefficients, are defined as functions of the particle concentration. As expected, no electrolyte parameters are required.

Despite being a very useful and widely used model, SPM does not capture the effects of the electrolyte. In order to incorporate these we need, at the very least, to consider the Single Particle model with electrolyte dynamics (SPMe).

\subsection{Single Particle Models with electrolyte dynamics (SPMe)}
As discussed in the previous section, the SPM performs very well at low C-rates but its accuracy decreases as the C-rate increases. Model accuracy can be improved by incorporating electrolyte dynamics, which gives rise to the next class of model, namely the Single Particle Model with electrolyte dynamics (SPMe). The main advantage of this model is that it retains most of the simplicity of the SPM, in the sense that it is still one-dimensional in space (albeit that three decoupled one-dimensional problems must be solved), but provides a significant improvement in performance. Recall that, as explained at the beginning of Section \ref{sec:DFN_to_SPM}, the SPMe refers to a family of models (of which \cite{BrosaPlanella2021,Marquis2019,Richardson2020} are particular instances) so the points discussed here apply to all the particular examples of SPMe.

The full details of this type of model can be found in Section \ref{sec:SPMe-type}, but the key idea is that in addition to the particle equations from the SPM, we also solve a PDE for the ion concentration in the electrolyte. Then, using the electrolyte concentration, we can calculate additional terms in the voltage expression to improve its accuracy. The complexity of the SPMe is slightly higher than that of the SPM. Apart from the PDEs for each particle, we now need to solve an additional quasi-linear PDE for the electrolyte ion concentration. However, the three spatially one-dimensional PDEs are decoupled so that they can be solved very efficiently. In this case, the model requires 30 parameters (apart from physical constants and the input, as shown in Table \ref{tab:parameters}): 9 for each electrode, 11 for the electrolyte, and the battery temperature. Similarly to the SPM, note that some of the electrolyte parameters, such as $D_\mre$ and $\sigma_\mre$, are also defined to be functions of the electrolyte concentration.

As shown in \cite{BrosaPlanella2021,Marquis2019,Richardson2020}, SPMe models can show very good agreement with the DFN model for a wide range of operating conditions, which makes them very useful in many practical applications. However, the SPMe does not capture the spatial distribution of lithium concentration across each electrode (only within a representative particle). In practice, as is demonstrated and explained in \cite{Richardson2020}, such spatial variations across the electrode are usually only significant for materials such as LiFePO$_4$, with flat open-circuit potential curves. For materials, such as NMC, with a significant gradient in their open-circuit potential curve the particles right across the electrode are all forced to maintain near identical surface lithium-ion concentrations, and therefore discharge at nearly the same rate. One notable weakness of the SPMe model occurs where the current becomes sufficiently large to cause electrolyte depeletion. In this scenario the agreement between the SPMe model and the DFN model breaks down. To capture these effects, which can play a very important role at high C-rates, especially where electrolyte depletion occurs or when incorporating degradation effects, we need to use the DFN model.

\subsection{Doyle-Fuller-Newman model (DFN)}
The Doyle-Fuller-Newman (DFN) model is the most ubiquitous physics-based model. Despite being significantly more complex than the SPM and the SPMe, this model is widely used because it captures the necessary features to represent the battery behaviour in an extremely broad range of operating conditions. Even though the DFN model is complex, it is still computationally affordable for many applications and, with accurate parameterisation, is capable of making extremely accurate predictions against experiment (see, for example, \cite{Chen2020,Ecker2015ii,Zulke2021}).

The key idea of the DFN model is to use an idealised geometry which still captures the multiscale nature of physics-based battery models. It assumes the cell has a one-dimensional planar geometry, with coordinate $x$, at the cell level (including the electrodes and separator) and a one-dimensional spherically symmetric geometry, with coordinate $r$, at the particle level. For this reason, this model is also known as the pseudo-two-dimensional (P2D) model. However, in contrast to the SPMe  where the problems in the $x$ and $r$ spatial coordinates are fully decoupled, in the DFN model there is full coupling between the problems solved in the $x$ and $r$ coordinates. This means that the DFN model is significantly more computationally complex than SPMe, being two-dimensional in space as opposed to SPMe which is only one-dimensional. Despite its relative computational complexity there are software packages capable of solving the DFN model extremely rapidly and efficiently \cite{Korotkin2021,Sulzer2021}, and while speed and computer resources are not particularly pressing issues for a single simulation of a discharge curve, they become much more significant when multiple coupled versions of the DFN model need to be solved, as for instance when modelling a thermally coupled pouch or cylindrical cell discharge.

Unlike the SPMe, which considers a single representative particle in each electrode, the DFN model includes electrode particles at all points across the width of the electrodes. Crucially, this allows the spatial distribution of lithium across the electrodes to be accurately modelled. This spatial distribution, which has been observed experimentally \cite{Thomas-Alyea2017,Yao2019}, plays a vital role when modelling a non-homogeneous interfacial current density which can lead to non-uniform degradation of the electrodes \cite{Yang2017}. The DFN model also captures the smoothing of the peaks in the $\dd V / \dd Q$ curves and electrolyte depletion, none of which are captured by the SPMe.

Numerical solution of the DFN model, using the method of lines, results in a system of differential algebraic equations (DAEs). In contrast, the solution of the SPM and SPMe (in its simplest form),  using the method of lines, results only in a system of ordinary differential equations (ODEs). Recall that, while numerical solution of a system of coupled ODEs (or coupled algebraic equations) is a relatively straightforward task, numerical solution of a system of DAEs can often result in an ill-conditioned problem and therefore more sophisticated numerical methods are required. While these sophisticated techniques do not present a problem for a standard desktop machine, especially if the code is efficiently implemented, they are typically too computationally expensive to be solved on the standard hardware used in battery management systems. Since the DFN model includes the same physics as the SPMe it requires the same number of parameters.

\begin{figure}
    \centering
    \includegraphics[width=\textwidth]{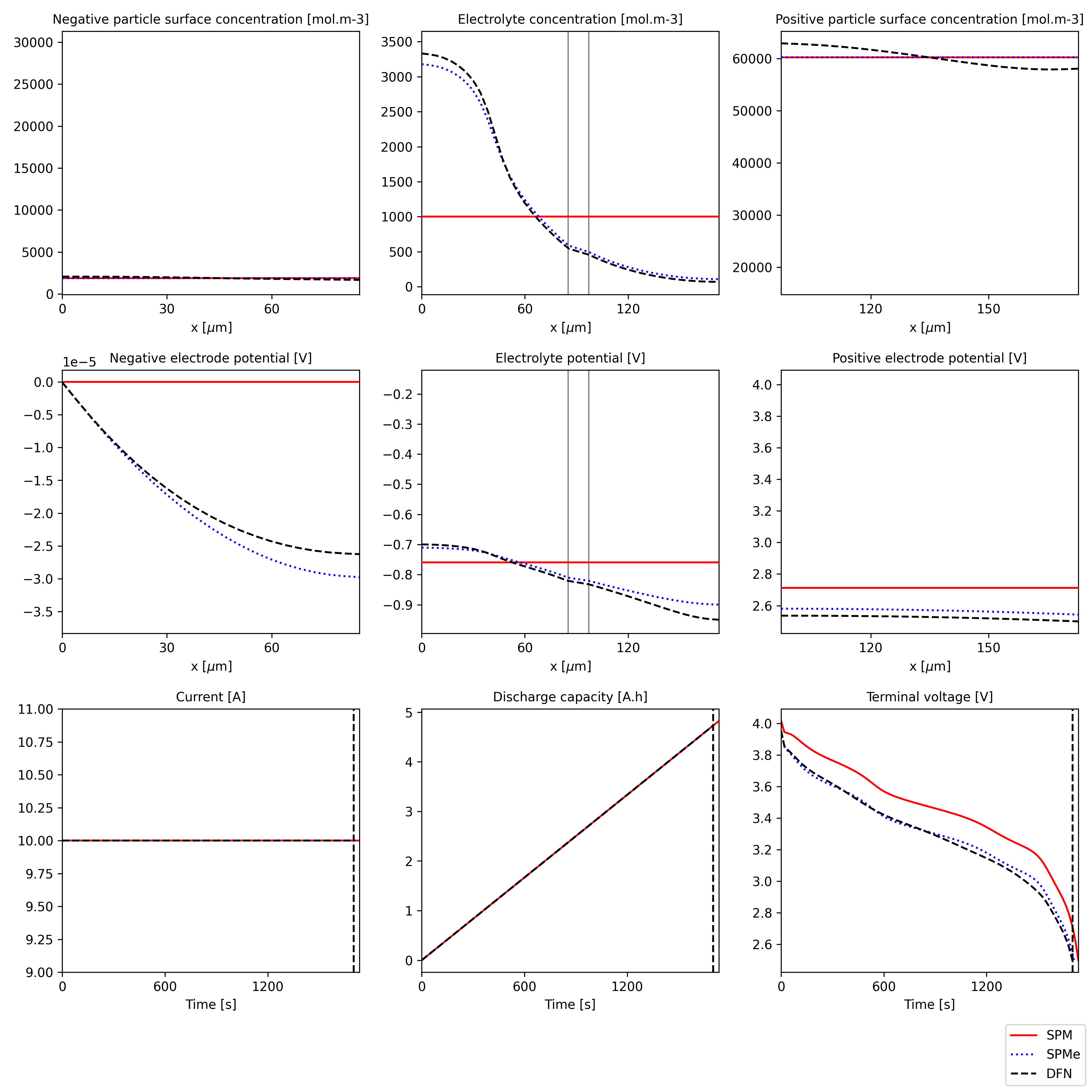}
    \caption{Comparison between the Single Particle Model (SPM), Single Particle Model with electrolyte (SPMe) and Doyle-Fuller-Newman model (DFN). The simulation corresponds to a 2C discharge with the parameters from \cite{Chen2020}. The snapshot corresponds to the end of discharge, for the interactive plot please see \url{https://github.com/FaradayInstitution/continuum-model-review}.}
    \label{fig:compare_internal_states}
\end{figure}

\begin{figure}
    \centering
    \includegraphics[width=\textwidth]{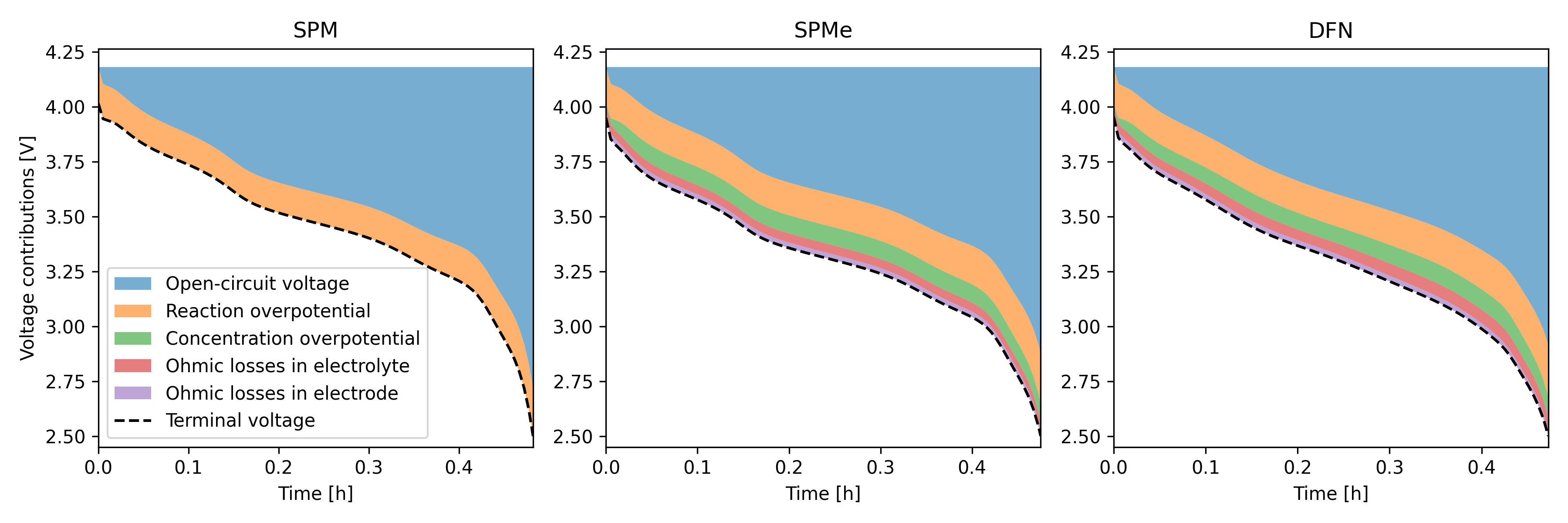}
    \caption{Comparison of the various contributions to the terminal voltage, as defined in \eqref{eq:V_SPMe_general}, for the Single Particle Model (SPM), Single Particle Model with electrolyte (SPMe) and Doyle-Fuller-Newman model (DFN). The simulation corresponds to a 2C discharge with the parameters from \cite{Chen2020}.}
    \label{fig:compare_voltage}
\end{figure}

At this point it is worth comparing the SPM, SPMe and DFN model, given that they all share the same simplified geometry which allows for direct comparison of their variables. The three models have been implemented in PyBaMM \cite{Sulzer2021}, and the code used can be found, jointly with some additional interactive examples, in the GitHub repository for this article (\url{https://github.com/FaradayInstitution/continuum-model-review}). Parameters are for a commercial battery with NMC 811 positive electrode and graphite-SiO$_{x}$ negative electrode (see \cite{Chen2020} for more details), and the simulations correspond to a 2C discharge. Figure \ref{fig:compare_internal_states} shows a comparison between the internal states (concentrations and potentials) and global states of the battery (current, discharge capacity and voltage). We observe that the SPMe model closely captures all the spatially distributed effects of DFN, but the SPM does not given that it does not include electrolyte dynamics. This results in turn in a poor prediction of the terminal voltage, especially at high C-rates. This becomes more clear when we compare the different contributions to the terminal voltage following the definition in \eqref{eq:V_SPMe_general}, as shown in Figure \ref{fig:compare_voltage}. We observe that the electrolyte contributions, split into the concentration overpotential and the ohmic losses in the electrolyte, are the main difference between the SPM and SPMe/DFN.

In conclusion, the DFN model is a very popular model and several extensions to account for additional physics have been proposed, as discussed in Section \ref{sec:extensions}. However, the DFN model only considers a one-dimensional problem at the cell scale and a one-dimensional problem at the particle scale, and therefore does not include any effects related to non-spherical particles or to the real three-dimensional geometry of the cell. In order to capture these effects, we need to use a more general homogenised model.

\subsection{Homogenised model}
Homogenised models can be regarded as an extension to the DFN model, and thus as a natural intermediate step between the DFN and microscale models. The key idea of the homogenised models is to generalise the DFN model to an arbitrary geometry, which could be up to 3+3D (that is, three-dimensions on both the cell and particle level), but to do so using an homogenised geometry for the porous electrodes. The advantage of this family of models is that we can resolve three-dimensional effects at the cell level (e.g. if we include the effect of current collectors \cite{Timms2021}) and we can also resolve three-dimensional effects in the particles (such as realistic particle geometries \cite{Kashkooli2016,Kashkooli2017,Kim2018}).

Despite having to cope with up to six spatial dimensions (up to three at each scale), this model is much simpler than a microscale three-dimensional model given that here we separate the two scales of the problem and thus we require a smaller resolution in each one. In order to simulate the model, apart from the parameters we need to provide the geometry of the cell and of a representative microstructure. This geometry replaces certain model parameters, such as the particle radii and the electrode thicknesses, as shown in Table \ref{tab:parameters}.

The homogenised model is significantly more complex than the DFN model because of the greater number of spatial dimensions and the more complicated geometry. In light of the geometrical complexity of the problem, especially at the particle level, it is usually discretised using the Finite Element Method (FEM) which can readily be adapted to such geometries \cite{Johnson2009}. In addition, certain limitations that could have been a cause for concern in the DFN model are critical here. The first issue is that of computing effective transport properties. In the DFN model the microstructure is assumed to consist of spherical particles, and therefore effective transport properties can be calculated, or estimated, from pre-existing results based on spherical packings (e.g. Bruggeman correlation \cite{Bruggeman1935,Tjaden2016} or Rayleigh method \cite{Bruna2015,Rayleigh1892}), but for the homogenised model these properties must be computed from imaging data or from idealised (albeit perhaps non-spherical) geometries. This leads us to the next issue which is: how strong is the separation of scales in a typical porous electrode geometry? Some methods to compute effective transport parameters are based on the assumption that there is a clear separation between the cell and particle length scales \cite{Hunt2020,LeHoux2020a}. However, for real batteries this separation of scales is not so clear (e.g. particle radius $\sim 5$ $\mu$m and electrode thickness $\sim 80$ $\mu$m, see \cite{Chen2020}) and, therefore, the estimated parameters might be questionable. Finally, another limitation of the homogenised model is that some of the new states introduced (such as the lithium distribution within particles, which now is non-symmetric) cannot be observed and therefore many of the outputs of the model cannot be validated.

Homogenised models allow us to include effects arising from the complex geometry of a real cell that cannot be treated satisfactorily with the DFN model, but they are still incapable of fully resolving a real geometry. To account for that, we need to consider the high-fidelity microscale model.

\subsection{Microscale model}
The high-fidelity microscale model is the most complex model of those considered in this review article and was first formulated and solved by Less et al. \cite{Less2012}. This model poses the equations for the physical laws described in Section \ref{sec:full_to_DFN} in a real geometry. This geometry captures the different materials and particles composing the porous electrode as well as the voids between which are filled by the electrolyte.

This model can be regarded as the most realistic model, as it has fewer underlying assumptions on the geometry or the operating conditions than any of the previous models. Additionally, it resolves each variable in the porous geometry so one can model their spatial distributions (e.g. to describe the ion flow around the particles), and any further reduction of this model can cause deviations in the predicted values \cite{Schmidt2020}. However, this does not mean it is the best model to use in practical applications because it poses two main challenges.

First, this model is extremely computationally expensive which means that it is not viable for most real applications \cite{Schmidt2020}. On top of that, the full microscale geometry of the battery is required (e.g. from tomography imaging \cite{Lu2020,Shearing2010,Shodiev2020}) which is also a complex task, especially when one considers that the extremely convoluted binder structure should be properly resolved. And second, as already mentioned for the homogenised model, many of the predictions of the spatial distributions of the quantities of interested cannot be validated as we currently do not have methods to observe them \textit{in operando}. Therefore, we have no way to check if the predictions of the behaviour of the internal states of the battery are even reasonable.

Of course, this model is still valuable from the theoretical point of view, and its importance will increase if in the future solving times are reduced (due to improvements in hardware and software) and new experimental techniques allow the monitoring of the internal states of an operating battery. But this should serve as a reminder that complexity for complexity's sake does not actually bring us closer to the main objective of battery modelling: to get a better understanding and more accurate predictions of the behaviour of these devices and, it is worth recalling, that extremely good agreement to experiment can be obtained from much less complex models, such as a properly parameterised DFN model \cite{Chen2020, Ecker2015ii,Zulke2021}.

\section{Coupled thermal-electrochemical models}\label{sec:thermal}

With the need for increasing power and faster charging capability any discussion of lithium-ion batteries will invariably involve a discussion on thermal management. Lithium-ion batteries are known to operate with peak efficiency at around 50$\degC$-60$\degC$ which corresponds to higher reaction rates and transport properties both in the electrodes and the electrolyte. However, they can easily reach temperatures far greater than this if not actively cooled. In extreme cases, under thermal or mechanical abuse, such as short circuiting the cells with a penetrating nail or via a lithium-metal dendrite through the separator, thermal runaway can occur due to the exothermic decomposition of the electrolyte at temperatures exceeding around 120$\degC$ \cite{Feng2018,Finegan2015}. Accurate prediction of internal temperatures is therefore critical for control systems, however it is complicated due to nonlinear heat generation and anisotropic thermal properties which also vary as a function of state-of-lithiation and temperature. Thermal conductivity in particular can vary by an order of magnitude between the in-plane/axial direction compared to the through-plane/radial direction of pouch/cylindrical batteries, respectively \cite{Taheri2013}. 

Many of the electrochemical processes occurring within a battery are strongly temperature dependent. Thus, in real devices, it is necessary to couple the electrochemical model of battery performance (e.g. the DFN model or Single Particle Model) to a thermal model in order to accurately capture its behaviour. This can greatly complicate the solution procedure because real devices, such as pouch or cylindrical batteries, are large enough to generate significant spatial temperature differences across their structure and, since the parameters in the electrochemical model are strongly temperature dependent, this requires that a version of the electrochemical model is solved at each point within the three-dimensional battery. From a computational perspective this presents a formidable challenge as can be seen by considering a thermally coupled DFN model. The battery is three-dimensional and large enough so that the temperature $T$ varies significantly across its structure, such that $T=T(x,y,z,t)$. Since many of the parameters appearing in the DFN model (in particular, diffusivities, reaction rates and the electrolyte conductivity) depend strongly on $T$ it is necessary to solve a version of the DFN model, with two spatial dimensions (i.e. a meso- and a micro-scopic dimension), at each macroscopic spatial point $(x,y,z)$ of the battery and couple this to a three-dimensional macroscopic thermal model. The resulting coupled electrochemical-thermal model has five spatial dimensions and leads to a problem with a high degree of computational complexity that requires significant computational resources to solve, in terms both of processor time and memory. Even when the DFN model is replaced by a Single Particle Model, the resulting coupled electrochemical-thermal model is still four-dimensional.

From a practical perspective, the solid state lithium diffusivity, the reaction rates and the electrolyte conductivity and diffusivity, have all been found to be well-described by an Arrhenius relationship with temperature. This means that their reference values taken at a particular reference temperature $T_\mathrm{ref}$ can be scaled using the relationship
\begin{equation}
    f(T) = f_\mathrm{ref} \exp\left( \frac{E_\mathrm{a}}{R} \left( \frac{1}{T_\mathrm{ref}}-\frac{1}{T} \right) \right),
\end{equation}
where $f$ is the parameter of interest, $f_\mathrm{ref}$ is the value of the parameter at the reference temperature $T_\mathrm{ref}$, $E_\mathrm{a}$ is the activation energy (per mole) and $R$ is the universal gas constant.
The open-circuit potentials also have a strong temperature dependence but this does not follow a straight-forward relationship and must be measured for each battery electrode chemistry.

In the literature, we find multiple approaches to include thermal transport and temperature effects into electrochemical models \cite{Bandhauer_2011,Bernardi_1985,Thomas_2003,Tranter2020b}, and there are already many reviews of the thermal management of lithium-ion batteries, including the various cooling systems \cite{An2017, Kim2019, Wang2016}. What follows here is a description of the different types of basic thermal model and a discussion of how they can be scaled up to tackle interesting and relevant problems faced by battery researchers which will prove useful to the continuum modeller seeking to extend their models or provide important information for practical applications, such as battery design or control.

\subsection{Coupling the electrochemical model to the thermal model}
Large format batteries (such as pouch, cylindrical and prismatic batteries) are formed by layering many extremely thin single cells (formed of a negative electrode, separator and positive electrode sandwiched between two current collectors) on top of each other, albeit in the case of cylindrical and prismatic batteries that there is only a single very long cell wound many times around a central core. The thickness of a single cell is so thin (usually no more than 200 $\mu$m) that it can be regarded as isothermal across its width (this result can be formally derived using asymptotic homogenisation \cite{Hennessy2020,Hunt2020}).

This means that at each point $(x,y,z)$ in the three-dimensional structure of a large format battery a single cell DFN problem should be solved in which the local temperature appears as a time-dependent parameter. From the solution of these local problems it is possible to deduce $\dot Q (x,y,z)$, the local rate of heating per unit volume. The volumetric rate of heating  $\dot Q=\dot Q_\mathrm{irr} + Q_\mathrm{rev}$ is the sum of the irreversible and reversible volumetric heating terms, $\dot Q_\mathrm{irr}$ and $Q_\mathrm{rev}$, respectively, which can both be computed from the local solution to the DFN model. It is noteworthy, however, that in the works published prior to 2021 that couple the DFN model to a thermal model, the formulae used for computing the irreversible heating term $\dot Q_\mathrm{irr}$ from the DFN model are, at best, only approximate. In particular they are inconsistent with the overall energy conservation law for the DFN model \eqref{econs}, as derived in \cite{Richardson2021heat} and discussed here in Section \ref{sec:econs}. The correct procedures for computing $\dot Q_\mathrm{irr}$ and $\dot Q_\mathrm{rev}$ from the solution to the DFN model are given in \cite{Richardson2021heat} and once established allow the local electrochemical (DFN) problems to be coupled to a macroscopic thermal problem for the temperature $T$ in a large format battery occupying the domain $\Omega_\mathrm{batt}$. This takes the form
\begin{align}
\rho c_p \pdv{T}{t} &= \nabla \cdot \left( \mathcal{K} \nabla T \right) + \dot Q_\mathrm{irr} + \dot Q_\mathrm{rev}, & \text{ in } \Omega_\mathrm{batt}, 
\end{align}
where $\rho$, $c_p$ and  $\mathcal{K}$ are the density, the specific heat capacity and the thermal conductivity of the battery, respectively. Note that large format batteries are typically highly anisotropic, because of their layered structure, which is why the thermal conductivity $\mathcal{K}$ is normally specified as a tensor. Appropriate initial and boundary conditions on this thermal problem are imposed. Boundary conditions strongly depend on the method by which the battery is cooled, typically either tab cooling or surface cooling. In the case of surface cooling typically the boundary condition is derived from Newton's Law of cooling and has the form
\begin{align}\label{mactherm}
-\mathcal{K} \nabla T \cdot \vb*{n} &= h (T-T_{\rm amb}), & \text{ at } \partial \Omega_\mathrm{batt},
\end{align}
where $\vb*{n}$ is the outward normal to the cooled surface ${\partial \Omega_\mathrm{batt}}$, $T_\mathrm{amb}$ is the exterior ambient temperature and $h$ is the effective cooling coefficient. In the case of tab cooling typically the temperature is specified where the tab joins the cell although it might be necessary to account for the thermal resistance of the tab.

It remains to determine appropriate (electrical) boundary conditions on the local DFN models. The simplest assumption to make is that the current collectors are equipotential, so that the local DFN models, no matter where they are located within the battery, experience the same potential difference. This is often appropriate for pouch batteries where the current path to the output tabs is relatively short but may not be so for cylindrical and prismatic batteries, which, because of their wound construction, give rise to much longer current paths to the output tabs than occur in a pouch battery. Physics-based models that account for the potential drop in the current collectors in pouch \cite{Timms2021} and cylindrical \cite{Tranter2020a} or prismatic \cite{Psaltis2022} batteries can be derived beginning from a three-dimensional macroscopic battery model, such as the DFN model, and can be simplified to give a range of reduced order models describing the electrochemical behaviour of large format batteries \cite{Marquis2020}. Since the problem of coupling the potential difference across the DFN elements to the current flow occurring in the current collectors adds a significant degree of complexity to the problem we do not discuss this further, other than to say that similar problems have been tackled in equivalent-circuit modelling of prismatic and cylindrical batteries (e.g. \cite{Li21}).

Before leaving this subject we note that it should be possible to start from a microscale thermal model of the constituent parts of the cell, for example that given by Latz and Zausch \cite{Latz2011}, and to homogenise this to obtain a macroscale thermal equation analogous to \eqref{mactherm}. However, such an analysis has not, as far as we are aware, been performed. We also note that the macroscale thermal equation can be coupled to single particle models but, while it is straightforward to compute the heating terms occurring on the right-hand side of \eqref{mactherm} for the simple SPM model, so far no-one has derived expressions for $\dot Q_\mathrm{irr}$ and $ \dot Q_\mathrm{rev}$ in the case of the SPMe-type models. This is because of the difficulty in accounting for electrolyte heating in these formalisms.

\subsection{Lumped thermal-electrochemical model}
The computational expense of accurately solving a fully coupled thermal-electrochemical model means that this task is beyond the reach of most solvers. Nevertheless thermal variations play a fundamental role in the behaviour of large format batteries. In the limit where heat dissipation to the environment is much slower than heat conduction within the battery, it can be shown that spatial variations in the temperature $T$ are negligible so that it can be approximated by the time-dependent function $T(t)$ (see \cite{BrosaPlanella2021}). At its simplest this results in coupling between a single DFN model, with parameters that vary with the temperature $T(t)$ and a lumped thermal model of the form
\begin{subequations}
\begin{equation}
    \theta \pdv{T}{t} = \frac{1}{\|\Omega_\mathrm{batt}\|} \int_{\Omega_\mathrm{batt}} \left(\dot Q_\mathrm{irr} + \dot Q_\mathrm{rev} \right) \dd V - h \frac{\|\partial \Omega_\mathrm{batt} \|}{\| \Omega_\mathrm{batt} \|}(T-T_\mathrm{amb}),
\end{equation}
where $\theta$ is the lumped volumetric heat capacity, and $\|\Omega_\mathrm{batt} \|$ and $\|\partial \Omega_\mathrm{batt} \|$ represent the volume of the battery and area of the cooling surface. This is solved subject to the initial condition 
\begin{equation}
    T(0) = T_0.
\end{equation}
\end{subequations}
This approach brings a significant simplification to the solution procedure if the resistance of the current collectors is negligible, so that all parts of the battery experience the same potential difference between the current collectors. Clearly, if this is not the case then we have to revert to solving a DFN model at each point in space, and the heating terms $\dot Q_\mathrm{irr}$ and $\dot Q_\mathrm{rev}$ are no longer spatially independent. In this scenario the resulting problem is almost as computationally expensive as solving the fully coupled model detailed above.

\section{Possible extensions}\label{sec:extensions}

The models discussed in Sections \ref{sec:full_to_DFN} and \ref{sec:DFN_to_SPM} are electrochemical models, which are the cornerstone of battery modelling. However, in many situations these models are not sufficient and we need to incorporate new physics. We have already discussed how to incorporate thermal models but these are not the only effects we can couple to electrochemical models. Coupling new physics to an electrochemical model leads to extensions of the models in Sections \ref{sec:full_to_DFN} and \ref{sec:DFN_to_SPM}. As with the pure electrochemical models, there are many examples in the literature where these models are posed \textit{ad hoc} for different levels of complexity (e.g. based on the DFN model and the SPM). However, this can easily lead to inconsistencies in the model, especially as the number of additional physical processes increases. The framework for model reduction presented in Sections \ref{sec:full_to_DFN} and \ref{sec:DFN_to_SPM} remains perfectly valid when new physics are added to the microscale model which, given its high resolution, is usually the best candidate to incorporate any additional physical phenomenon. In this section we present possible extensions to the electrochemical model. We do not intend it to be an exhaustive list with all the details of each model, as each family would deserve its own review article. Instead, we aim for this section to be a starting point, which can serve as an introduction to these extensions and that points the reader to the relevant references. Note that each subsection introduces a significant amount of new notation, so each subsection is self-contained and this notation is not included in \ref{sec:nomenclature}.

\subsection{Phase-field models}

Physical and electrochemical phenomena occurring in lithium-ion batteries are complex and nonlinear. A good example of this is the ion intercalation dynamics in the solid particles, which involves lithium diffusion accompanied, in most cases, by phase transformations in the electrode material (for example lithium iron phosphate as a positive electrode material involving micrometer-sized particles \cite{Singh2008}). The models discussed in Sections \ref{sec:full_to_DFN} and \ref{sec:DFN_to_SPM} assume that diffusion drives the transport of intercalated lithium (e.g. \eqref{eq:DFN_electrode_diffusion} for the DFN model and \eqref{eq:SPMe2_particle} for SPM-type models). This approach either neglects any phase separation or tries to mimic the phase transformation process by introducing a highly nonlinear effective diffusion coefficient, which strongly depends on local lithium concentration and rapidly fluctuates in a wide interval (e.g. diffusion coefficient measurements vary between $10^{-11}$ and $10^{-8}$ cm$^2$/s in \cite{Ecker2015i}). However, it should be noted that a nonlinear diffusion model is unable to capture the hysteresis in the open circuit voltage that is observed for certain electrode materials as they are slowly charged and then discharged.

An alternative to modelling lithium transport in the electrode materials using a nonlinear diffusion equation is to use the Cahn–Hilliard reaction (CHR) theory that naturally captures the phase transformation \cite{Cahn1958i,Singh2008}. The CHR model is a special case of the phase transformations model developed by Bazant \cite{Bazant2013} based on non-equilibrium thermodynamics. It corresponds to rate limitation by diffusion and describes bulk phase separation driven by heterogeneous reactions localised on the surface and described by a flux matching boundary condition. In an arbitrary domain $\Omega$ it reads
\begin{subequations}
\begin{align}\label{eq:phase_field_cahn_hilliard}
\pdv{c}{t} &= -\nabla \cdot \mathbf{F}, & \text{ in } \Omega,\\
-\mathbf{F} \cdot \vb*{n} &= R \left( \left\{ \frac{\delta G}{\delta c} \right\} \right), & \text{ at } \partial \Omega,
\end{align}
\end{subequations}
with
\begin{align}
\mathbf{F} &= - M c \nabla \frac{\delta G}{\delta c},
\end{align}
where $c$ is the concentration of lithium, $M$ is the mobility, $R$ is the volumetric reaction rate, and $G$ is the total free energy.

The particular case of the CHR model with Butler–Volmer kinetics for a spherically symmetric, isotropic, strain-free, electron-conducting particle of radius $R_k$ for $\kin{n,p}$, as shown in \cite{Zeng2014}, can be written in the form
\begin{subequations}\label{eq:phase_field_particle}
\begin{align}
\pdv{c_{k}}{t} &= - \frac{1}{r^2} \pdv{}{r} \left( r^2 F_k \right), & \text{ in } 0 < r < R_k,\\
F_k &= 0, & \text{ at } r=0,\\
F_k &= \frac{j_k}{F}, & \text{ at } r=R_k,\\
c_k &= c_{k0}, & \text{ at } t=0.
\end{align}
with
\begin{align}\label{eq:phase_field_particle_F}
F_k &= - \frac{D_{k0}}{k_B T} c_{k} (c_k^{\max} - c_{k}) \pdv{\mu_k}{r},
\end{align}
\end{subequations}
where $c_k$ is the lithium concentration, $j_k$ is the Butler-Volmer surface current, $c_{k0}$ is the initial concentration and $c_k^{\max}$ is the maximum concentration.

The parameter $D_{k0}$ in \eqref{eq:phase_field_particle_F} is the dilute-solution limit that can be expressed via mobility $M_k$ using the Einstein relation for the tracer diffusivity as
\begin{align}\label{eq:phase_field_particle_mobility}
D = M_k k_B T = D_{k0} \left( 1 - \frac{c_k}{c_k^{\max}}\right).
\end{align}
The chemical potential $\mu_k$ in \eqref{eq:phase_field_particle_F} can be roughly approximated in terms of the open-circuit potentials $U_k$ in the corresponding domain $\kin{n,p}$ as $\mu_k = -F U_{k}$, where $F$ is the Faraday constant. For the Cahn–Hilliard regular solution model \cite{Cahn1959ii,Cahn1958i,Cahn1959iii}, the chemical potential can be represented as
\begin{align}\label{eq:phase_field_chem_potential}
\mu_k = k_B T \ln \left( \frac{c_k}{c_k^{\max} - c_k} \right) + H \left(\frac{c_k^{\max} - 2c_k}{c_k^{\max}} \right) - \frac{\omega}{(c_k^{\max})^2} \nabla^2 c_k,
\end{align}
where $k_B$ is Boltzmann constant, $T$ is the absolute temperature, $H$ is the enthalpy of mixing, and $\omega$ is the gradient energy penalty coefficient.

For the chemical potential $\mu_k$ written in the form \eqref{eq:phase_field_chem_potential} we need two additional boundary conditions for the concentration $c_k$:
\begin{align}
\pdv{c_{k}}{r} &= 0, & \text{ at } r=0,\\
\pdv{c_{k}}{r} &= \beta_k, & \text{ at } r=R_k.
\label{eq:phase_field_additional_bcs}
\end{align}
Here $\beta_k$ is the (constant) concentration gradient at the particle surface which is related to the surface-to-bulk phase boundary tension ratio or, according to Young's law, to the equilibrium contact angle $\theta_k$ as $\beta_k = c_k^{\max} \cos{(\theta_k)} / \lambda_k$, where $\lambda_k$ is the phase boundary thickness \cite{Zeng2014}. Equations  \eqref{eq:phase_field_particle}-\eqref{eq:phase_field_additional_bcs} can be used to replace the lithium diffusion equation employed by the classic DFN model. They naturally describe phase transformation phenomena but on the other hand, the model involves a set of additional unknown variables (e.g. chemical potentials) that are difficult to estimate and may require computationally expensive micro-scale simulations.

\subsection{Capacitance layers}

The models developed so far include a jump in potential at the interface between the solid phase and electrolyte phase.
In reality, the potential changes continuously in a nanometer-thick layer known as the double layer.
Even though it could be modelled as a distributed effect, the layer is so thin that it is more convenient to model the capacitance as an interfacial effect, including an additional current due to the changing potential difference.
This capacitance term is often ignored (as in Sections \ref{sec:full_to_DFN} and \ref{sec:DFN_to_SPM}) since the time scale of capacitance effects, a few seconds, is much shorter than the time scale of hours to years that is usually considered in battery models.
However, in some scenarios that occur over short time scales, such as extreme fast charging, capacitance effects could play an important role.

Furthermore, including the double layer current in the model can also be useful for mathematical reasons, since it converts one of the algebraic equations into a differential equation and hence improves the conditioning of the model.
In fact, in the homogenised one-dimensional system, the DFN model can be transformed to a system of parabolic PDEs which, when semi-discretised in space, becomes a system of ODEs and hence can be solved using a wider range of algorithms than the ones required for the DAE system resulting from the semi-discretisation of the standard DFN model (which consists of parabolic and elliptic PDEs).

At the microscale, we model the capacitance as an additional current at the interface between the solid and electrolyte phases \cite{Sulzer2019Thesis}, so that \eqref{eq:microscale_internal_bc} becomes
\begin{subequations}\label{eq:microscale_internal_bc_capacitance}
\begin{align}
     F \tilde{\vb*{N}}_k \cdot \vb*{n}_k &= \tilde{\vb*{i}}_k \cdot \vb*{n}_k = j\dl_k + \tilde{j}_k, & \text{ at } \vb*{x} \in \partial \Omega_k^\mathrm{in},\\
     F \tilde{\vb*{N}}_k \cdot \vb*{n}_\mre &= \tilde{\vb*{i}}_\mre \cdot \vb*{n}_k = -(j\dl_k + \tilde{j}_k), & \text{ at } \vb*{x} \in \partial \Omega_\mre^\mathrm{in}.
\end{align}
\end{subequations}
The double-layer current depends linearly on the rate of change of the potential difference,
\begin{equation}\label{eq:capacitance}
    j\dl_k = C\dl\pdv{}{t}\left(\phi_k - \phi_\mre\right),
\end{equation}
where the double-layer capacitance, $C\dl$, is usually taken to be a constant.
It may be more physically accurate to allow the double-layer capacitance to be a function of the potentials, in which case it would move into the time derivative in \eqref{eq:capacitance}, but since the capacitance time scale is so small this is of little importance in practice.
While \eqref{eq:capacitance} is a model of an ideal capacitor, it is also possible to model the double-layer as a non-ideal capacitor \cite{Chu2019}.

The capacitance terms propagate trivially through the homogenisation process, so that equations (\ref{eq:homogenised_particle}b), (\ref{eq:homogenised_electrode}a), (\ref{eq:homogenised_electrolyte}a) and (\ref{eq:homogenised_electrolyte}b) are replaced, respectively, by
\begin{subequations}
\begin{align}
    F \tilde{\vb*{N}}_k \cdot \vb*{n}_{k} &= j\dl_k + \tilde{j}_k, 
    && \text{ at } \vb*{X} \in \partial \Omega_{k}^\mathrm{in},\\
    \nabla \cdot \vb*{i}_k &= - b_k \left(j\dl_k + j_k\right), 
    && \text{ in } \vb*{x} \in \Omega_{k},\\
    \varepsilon \pdv{c_\mre}{t} + \nabla \cdot \vb*{N}_\mre &= \frac{b\left(j\dl + j\right)}{F}
    && \text { in } \vb*{x} \in \Omega,\\
    \nabla \cdot \vb*{i}_\mre &= b \left(j\dl + j\right), 
    && \text { in } \vb*{x} \in \Omega.
\end{align}
\end{subequations}

In a one-dimensional setting with capacitance effects included, the Doyle-Fuller-Newman model can be converted into a system of parabolic PDEs as follows \cite{Sulzer2019FasterPhysical}. First, we define the surface potential difference
\begin{equation}\label{eq:Deltaphi}
    \Delta \phi_k = \phi_k - \phi_\mre, \quad \text{ in } \vb*{x} \in \Omega_{k}.
\end{equation}
Substituting \eqref{eq:Deltaphi} into \eqref{eq:DFN_phik} and  \eqref{eq:DFN_phie}, and using the fact that for one-dimensional problems $i_\mre + i_k = i_\mathrm{app}$, we obtain a system of simultaneous equations for the electrolyte current,
\begin{subequations}\label{eq:is_ie_capacitance}
\begin{align}
    i_\mathrm{app} - i_{\mre,k} &= -\sigma_k\pdv{}{x}\left(\Delta \phi_k + \phi_{\mre,k}\right),
    \\
    i_{\mre,k} &= \sigma_\mre (c_\mre) \mathcal{B}(x) \left( -  \pdv{\phi_{\mre,k}}{x} + 2 (1-t^+) \frac{R T}{F} \pdv{\log c_\mre}{x} \right),
\end{align}
\end{subequations}
where $i_{\mre,k}$ denotes the electrolyte current in the subdomain $k$.

We can then eliminate $\partial \phi_{\mre,k}/\partial x$ from \eqref{eq:is_ie_capacitance} to write $i_{\mre,k}$ as a functional of $\Delta \phi_k$,
\begin{equation}
    i_{\mre,k} = \frac
        {\pdv{\Delta \phi_k}{x} + 2 (1-t^+) \frac{R T}{F} \pdv{\log c_\mre}{x} + i_\mathrm{app} / \sigma_k}
        {1 / \left(\sigma_\mre (c_\mre) \mathcal{B}(x)\right) + 1 / \sigma_k},  
        \quad \text{ in } \vb*{x} \in \Omega_{k}.
\end{equation}
Thus the two elliptic equations \eqref{eq:DFN_phie} and \eqref{eq:DFN_phik} can be replaced by a single parabolic equation for $\Delta \phi_k$,
\begin{equation}
    \pdv{\Delta \phi_k}{t} = \frac{1}{b_kC\dl}\left(\pdv{i_{\mre,k}}{x} - b_kj_k\right),  \quad \text{ in } \vb*{x} \in \Omega_{k},
\end{equation}
with initial and boundary conditions following trivially from linear combinations of the initial and boundary conditions in the DFN model.
When semi-discretised with the method of lines, the capacitance-DFN model becomes a system of ODEs, and hence can be solved by simpler numerical methods than the standard DFN model.

Finally, having found $i_\mre$, the electrolyte potential is found by integrating \eqref{eq:DFN_phie},
\begin{equation}
    \phi_\mre = \int_{0}^{x} \left( 2 (1-t^+) \frac{R T}{F} \pdv{\log c_\mre}{x} - \frac{i_{\mre,k}}{\sigma_\mre (c_\mre) \mathcal{B}(s)} \right) \dd s 
    - \left.\Delta \phi_k\right\rvert_{x=0}, 
\end{equation}
and the electrode potential (and hence voltage) by applying \eqref{eq:Deltaphi}.
Note that this manipulation is only possible in one-dimensional problems, since $i_\mre + i_k \neq i_\mathrm{app}$ in higher dimensions \cite{Sulzer2019Thesis}.

\subsection{Multi-particle electrodes and graded electrodes}

The standard DFN model (and hence the associated reduced order models) assumes that, within an electrode, all particles are spheres with the same radius. In reality, at the microscale, all the particles have different shapes and sizes. While retaining the spherical assumption, the different particle sizes can be incorporated into the DFN model in two different ways: taking the particle size to be a function of $x$ (graded electrodes), and including a distribution of particle sizes at each point in $x$ (particle-size distribution).
It is also (in theory) possible to combine both of these approaches by making the particle size distribution itself be graded.

\paragraph{Graded electrodes}
In graded electrode models, the particle radii themselves are taken to be a function of $x$, $R_k(x)$.
A na\"ive numerical implementation would require a different mesh for each particle equation. This can be avoided by first rescaling $r=\hat{r}R_k(x)$, so that \eqref{eq:DFN_electrode_diffusion} becomes
\begin{align}\label{eq:DFN_electrode_diffusion_graded_electrode}
    \pdv{c_k}{t} &= -\frac{1}{\hat{r}^2R_k(x)}\pdv{}{\hat{r}}\left( \hat{r}^2 N_k \right), & N_k &= -\frac{D_k (c_k)}{R_k(x)} \pdv{c_k}{\hat{r}},
\end{align}
with $0 \leq \hat{r} \leq 1$ (note that $R_k(x)$ is independent of $\hat{r}$ and thus commutes with the $\hat{r}$-derivative).

Since the Single Particle Model consists of a representative particle in each electrode, graded electrodes cannot be included in that model, nor in the SPMe of Marquis et al.~\cite{Marquis2019} for the same reason.
However, the corrected SPM introduced by Richardson et al.~\cite{Richardson2020} does allow for inclusion of graded electrodes, and thus could be used when a reduced order model that includes graded electrodes is required.

\paragraph{Particle-size distributions}
In particle-size distribution models, there is a distribution of particle sizes at each point in $x$; specifically, a distribution of active material volume fractions, $\varepsilon^\theta_k$, occupied by particles of size $R^\theta_k$.
Hence the particle concentrations are now pseudo-3D, $c_k = c_k(x,r,\theta)$, where $\theta$ parameterises the `particle size' dimension.
As a result, the interfacial current density is a function of $x$ and $\theta$,
\begin{equation}
    j^\theta_k(x,\theta) = j_{k0}(c_\mre(x),c_k(x,R_k,\theta)) \sinh \left[ \frac{F}{2 R T} \left(\phi_k(x) - \phi_\mre(x) - U_k(c_k(x,R_k,\theta))\right) \right].
\end{equation}
Equation~\eqref{eq:DFN_particle} is unchanged but is now parameterised by $\theta$ as well as $x$.
The total interfacial current density, which feeds into the macroscale equations for electrolyte concentration, electrolyte potential, and solid-phase potential, is given by
\begin{equation}\label{eq:particle_size_integral}
    b_kj_k = \int b^\theta_k j^\theta_k \dd \theta,
\end{equation}
where $b^\theta_k = 3\varepsilon^\theta_k/R^\theta_k$ is also a function of $\theta$ since both $\varepsilon^\theta_k$ and $R^\theta_k$ are.

As indicated in \eqref{eq:particle_size_integral}, the solid phase and electrolyte potentials and the electrolyte concentration are the same at the surface of each particle at a particular point in $x$, but the surface concentrations $c_k$ (and therefore $j_{k0}$ and $U_k$) depend on $\theta$.
This is because lithium can diffuse from the centre to the surface of the particle more rapidly in smaller particles than in larger ones.
Typically, \eqref{eq:particle_size_integral} is solved by discretising the particle-size distribution so that the integral becomes a sum of a finite number of particle sizes.

When considering this level of detail, one should bear in mind that the separation of scales between the particle radius ($\sim5$ $\mu$m) and the electrode thickness ($\sim80$ $\mu$m) is not very clear (values taken from \cite{Chen2020}), which limits the accuracy of the homogenisation of the particle-size distribution.
Therefore, it may be sensible to consider \textit{either} particles of uniform size but with varying concentrations in the through-cell dimension (i.e. the standard DFN model) \textit{or} a particle-size distribution within an SPM-like model, for example as proposed by Kirk et al.~\cite{Kirk2020} and Mohtat et al.~\cite{Mohtat2020}.

\subsection{Mechanical models}

Although lithium-ion batteries are devices based on electrochemical reactions, they also face coupled mechanical problems, such as volume expansion that leads to stress generation and particle cracking, which contribute to capacity loss \cite{Palacin2016}. The volume change of active materials under (de)lithiation during battery cycling originates from the varying atomic layer distance with varying lithium content in the active materials. Graphite, for instance, expands approximately 10 vol\% from $\text{C}_6$ to $\text{LiC}_6$ \cite{Koerver2018,Popp2020}, with the volume change of other active materials summarised in \cite{Popp2020}. These stresses and cracks in electrode particles bring strong nonlinearities to battery (dis)charge responses, which challenge current battery models as shown in various reviews \cite{Zhao2016,Zhao2019}. 

Early contributions to mechanical models for lithium-ion batteries stem from Christensen and Newman \cite{Christensen2006} and Zhang et al. \cite{Zhang2007}. These models coupled the mechanics and electrochemistry in lithium-ion batteries and describe the volume change and stresses in electrode particles as a function of lithium concentration. The magnitude of radial ($\sigma_r$) and tangential ($\sigma_\theta$) stresses in electrode particles $r \in [0, R]$ are determined by 
\begin{subequations}
\begin{align}
    \sigma_r(r) &= \frac{2\Omega E}{3(1-\nu)} \left(\frac{1}{R^3}\int_{0}^{R} \hat{c}r^2\dd r -\frac{1}{r^3} \int_0^r \hat{c} \hat{r}^2 \dd \hat r \right),\\
    \sigma_\theta(r) &= \frac{\Omega E}{3(1-\nu)} \left(\frac{2}{R^3}\int_{0}^{R} \hat{c}r^2\dd r -\frac{1}{r^3} \int_0^r\hat{c} \hat{r}^2 \dd \hat r - \hat{c} \right),
\end{align}
\end{subequations}
where $\Omega$ is the partial molar volume, $\hat{c}=c-c_0$ where $c_0$ is a reference concentration, $E$ is Young's modulus, $\nu$ is Poisson's ratio, and $R$ is the particle radius. Within the academic literature, $\Omega$ has been assumed to be either a constant value or a function of concentration. Here, the lithium concentration gradient is a key driving factor of the magnitude of the particle stresses. However, these stresses have also been shown to influence the solid-phase diffusion, which can be described by
\begin{subequations}
\begin{align}
    \pdv{c}{t} &= D \left[\pdv[2]{c}{r} + \frac{2}{r} \pdv{c}{r} + \theta_M \left( \pdv{c}{r} \right)^2 + \theta_M (c-c_0) \left( \pdv[2]{c}{r} + \frac{2}{r} \pdv{c}{r} \right) \right], & \text{ in } 0 < r < R,
\end{align}
\begin{align}
    \pdv{c}{r} &= 0, & \text{ at } r = 0,\\
    (1+\theta_M (c-c_0)) D \pdv{c}{r} &= - \frac{j}{F}, & \text{ at } r = R.
\end{align}
\end{subequations}
where $\theta_M=\frac{\Omega}{RT}\frac{2\Omega E}{9(1-\nu)}$, $D$ is the diffusion coefficient, $j$ is the Butler-Volmer current density, and $F$ is the Faraday constant. This equation can be used to substitute the particle equations in the DFN model.

The mechanical model has been applied to SPMe to model battery capacity degradation \cite{Li2018a} and to the DFN model to study the non-uniform stress distribution across the two electrodes \cite{Ai2020}, finding that large stresses appear closer to the separator leading to local particle cracking. These cracks release new surfaces for the SEI formation and growth, which consume additional accessible lithium ions and accelerate the battery capacity degradation \cite{Li2018a}. Many methods have been developed to control the level of stresses, which can be classified into two main groups: material design and optimised charging profiles. The latter is often preferred because of its low cost. By varying the charging profiles this approach can also achieve longer lifespan of lithium-ion batteries (e.g. reducing the charging current at high states-of-charge) although this benefit can vary markedly for battery cells with different chemistry \cite{Keil2016}. Optimal charging profiles can be obtained using the mechanical model by decreasing the charging currents when the stress reaches a critical threshold \cite{Suthar2014}.

However, despite all this good progress, the developed models use assumptions which limit their applications to more complex working conditions (e.g. spherical particles, homogeneous material properties, small deformation...). Topics of stress affected diffusion \cite{Zhang2007}, large deformation \cite{Li2016}, anisotropy \cite{Liu2011} and microstructure \cite{Mendoza2016} are also important to capture, and more details can be found in the references provided.  

\subsection{Degradation models}

Whilst accurately predicting battery performance at the beginning of its life is useful, prediction of future lifetime is perhaps even more important. This is often a complex task due to the diversity of battery degradation modes and the interdependency between them, leading to a path dependent behaviour \cite{Raj2020}. A summary of these mechanisms can be found in the review by Edge et al. \cite{Edge2021} and the details on how to model them can be found in \cite{OKane2022}. Here, we review the most relevant modes from a modelling perspective.  

\subsubsection{Negative Electrode Degradation mechanisms}

\paragraph{Solid electrolyte interphase (SEI) layer growth}
The SEI is a passivation layer that forms on most negative electrode materials due to the thermodynamic instability of the solid-electrolyte interface, and is often cited as one of the main degradation modes. Several DFN models \cite{Ekstrom2015,Ning2006,Ramadass2004,Safari2009} of SEI formation exist in the literature to predict the capacity fade, with the pioneering work of Safari et al.~\cite{Safari2009} summarising the SEI formation layer model well with clear physical meaning. In this work, solvent diffusion through the SEI layer was modelled, followed by electrochemical reactions governed by the Tafel equation, including both kinetic and diffusion limitations. The growth of the SEI layer results in an increased resistance and capacity fade through loss of lithium inventory. Important input parameters of SEI models are OCP, electrochemical reaction rates, diffusion coefficients, and resistance. 

\paragraph{Lithium plating}
Lithium plating is a side reaction where metallic lithium deposits on the negative electrode surface instead of intercalating into it. Lithium plating and stripping (the inverse process of plating) can be modelled by including an additional Butler-Volmer equation for the side reaction in the DFN model, where the general assumption is that plating occurs when the negative electrode potential compared to a lithium reference electrode is less than 0 V. This basic lithium plating model was first proposed by Arora, Doyle and White \cite{Arora1999}, but the most comprehensive model was proposed by Yang et al.~\cite{Yang2018b} by considering dependence on both electrolyte and plated lithium concentrations. In a recent study, O’Kane et al.~\cite{OKane2020} incorporated a nonlinear diffusion model into Yang’s model to account for phase transitions in the graphite and showed that nonlinear diffusion plays an important role to yield qualitatively different final results.

\paragraph{Particle fracture}
Particle fracture is caused by the substantial volume change and the resulting stress of electrode materials during cycling, especially for particles with large sizes or under high C-rates \cite{Christensen2006b}. Particle fracture leads to battery degradation, with effects such as binder detachment \cite{Foster2017}, loss of active materials \cite{Reniers2019} and loss of lithium inventory by additional SEI growth on new surfaces of cracks \cite{Deshpande2012}. The modelling of particle fracture can be through empirical functions to account for the new surfaces of cracks for the additional SEI growth \cite{Ekstrom2015}, but this kind of models is only accurate for moderate C-rates. Physics-based models have been developed using the fatigue crack model of Paris' law \cite{Deshpande2012}, which shows better accuracy for battery cells at high C-rates. There are also three-dimensional studies on the microstructure of electrode materials \cite{Xu2019}, using finite element methods, but they are computationally expensive and usually limited to beginning of life simulations.

\subsubsection{Positive Electrode Degradation Mechanisms}

\paragraph{Phase Change and Oxygen Evolution}
Oxygen evolution from the positive electrode has been conventionally modelled as oxidation of electrolytes at the positive electrode side using a simple kinetically limited Tafel equation, with this kind of model being proposed in several literature reviews \cite{Lin2013,Reniers2019}. Jana et al.~\cite{Jana2019} proposed that the capacity fade is a linear function of the oxidation current density, which they used in the Tafel equation to model the electrolyte oxidation at the positive electrode. This oxidation reaction also produces H$_2$ which in turn enhances the transition metal dissolution. However, theoretical understanding and modelling for the source of the oxygen evolution and its effect on the capacity fade is lacking. In this regard, Ghosh et al.~\cite{Ghosh2020} proposed a shrinking core model based on thermally coupled SPM in which lattice oxygen release is considered due to the phase transition of the positive electrode materials at higher states-of-charge and the degradation mechanism is controlled by either diffusion of lattice oxygen through the phase transformed surface layers or the reaction kinetics at the bulk/surface layer interface of the electrode.

\paragraph{Transition Metal (TM) Dissolution and positive electrode solid electrolyte interphase (pSEI) formation}
TM dissolution happens at the electrode-electrolyte interface due to chemical side reactions which dissolve TMs and form pSEI. TM dissolution at the positive electrode is modelled using a first order chemical reaction, limited by concentration of H$^+$ ions in the electrolyte \cite{Dai2013}. H$^+$ ions are generated from LiPF$_6$ salt dissociation in the electrolyte and solvent oxidation at the positive electrode. While LiPF$_6$ dissociation in the presence of H$_2$O is modelled using a chemical reaction rate, solvent oxidation is modelled using Tafel kinetics \cite{Dai2013}. Lin et al.~\cite{Lin2013} provided detailed DFN model equations for TM dissolution at the positive electrode coupled with SEI layer formation at negative electrode. The TM deposition on the negative electrode is also included in the model. The growth of the pSEI can be modelled similarly to any of the SEI layer growth models. Lack of agreed equations in the literature make modelling of the positive electrode degradation mechanisms still an active area of research.

\section{Conclusions}
The purpose of this work has been to review cell scale continuum electrochemical models for lithium-ion batteries of varying degrees of complexity. We started by considering an extremely detailed microscopic model that is capable of simulating all of the individual components that make up the cell, down to the scale of individual electrode particles. We then applied homogenisation techniques to the microscale model to obtain the homogenised model. This model introduces the concept of the multi-scale model: the electrode and electrolyte equations are defined in a simplified domain (and use effective transport properties to capture the effects of the porous geometry) and they are coupled with a microscale problem that resolves lithium transport in the particles. Next we reviewed the Doyle-Fuller-Newman (DFN) model, which can be seen as a particular case of the homogenised model for a very simple geometry: one-dimensional planar geometry in the electrodes and one-dimensional spherically symmetric geometry for the particles. The homogenisation combined with the simple geometry yield a model that is tractable and can be easily solved on a laptop. However, some applications require simpler (i.e. faster to solve) models, so in Section \ref{sec:DFN_to_SPM} we presented some simplifications of the DFN model: the {\it so-called} Single Particle Models. These represent a considerable reduction in computational complexity when compared to the DFN model and, while they still necessitate that macroscopic equations be solved for the current flows and the lithium diffusion in the electrolyte, they only require that lithium transport equations are solved in one (or at worst a small number of) electrode particle in each electrode. In order to aid in the selection of a model, appropriate to the requirements of the user, the main advantages and disadvantages of these three approaches are briefly outlined below (full details can be found in Section \ref{sec:comparison}).

The full benefit of employing the detailed microscopic model can only be realised if it used in conjunction with accurate three-dimensional microscopic data of the cell (e.g. \cite{Pietsch2017}), resolved down to the level of individual electrode particles, in order to generate the geometry over which the model is solved. It may be necessary to resolve the crystal structures of the electrode particles, which for certain highly anisotropic materials, such as graphite, can heavily influence the transport properties of the electrode particles. Moreover, given the very thin chain like structures typically formed by the conductive additives in the binder domain, it is almost impossible to obtain a properly resolved microscopic description of conduction within the electrode. Full microscopic simulations are not only extremely expensive to set up, in terms of domain specification, they are also, because of their computational complexity, very costly to run. However, in spite of these disadvantages, microscopic models are still probably the best way of assessing the effects that the microstructure (and changes to it) has on the battery performance. The homogenised model, even though significantly simpler, also poses the geometry characterisation challenge, with the additional constraint that this geometry needs to be a periodic structure that is representative of the whole electrode. In contrast, the DFN model is significantly less computationally complex than the full microscopic model and yields model equations that are two-dimensional in space (one microscopic dimension in the electrode particles and one mesoscopic dimension across the cell) and that can, if properly parameterised, accurately predict experiments. Furthermore, the geometry on which these equations are solved is easy to specify, albeit that a number of the parameters have to be computed (or at least estimated) from the microstructure of the electrodes. 

The final class of models that we reviewed were the Single Particle Models. Within this class, we distinguished between Single Particle Models with electrolyte dynamics (SPMe) and without (SPM). Note that even within each type there are multiple formulations of these models, but they are all very similar. These models are less computationally complex than the DFN model, typically involving one fewer spatial dimension than required for the DFN model and, for most electrode chemistries (except those with flat open-circuit potentials, such as lithium iron phosphate), they can be formally derived as a limit of the DFN model. Where applicable, they can give extremely good agreement with the DFN model, even at quite high rates of charge and discharge, although they cannot accurately describe scenarios in which electrolyte depletion occurs and so are unsuitable for very high rates. The reduction in computational complexity of the single particle models compared to the DFN model gives rise to significant advantages in computationally intensive applications, such as parameter estimation and optimal cell design, and indeed have been used to conduct the thermally coupled simulation of a cylindrical battery. SPM also has the notable advantage of having fewer parameters than the DFN model and so is more straightforward to fit to data.

After reviewing the electrochemical models, we introduced coupled thermal-electrochemical models for large format batteries (such as pouch, cylindrical and prismatic batteries). These models are necessary to describe how a battery heats up during operation, which has a huge impact on its behaviour. We finished by providing a brief overview of possible extensions to the electrochemical models, including: phase-field models, capacitance layers, multi-particle and graded electrodes, mechanical effects and degradation. All these models need to be properly coupled to an electrochemical model, and the derivation of simplified models accounting for these additional physics should be done using the same framework as we introduced for electrochemical models.

These model extensions are connected to the main areas for further research in the field of continuum models for lithium-ion batteries. The principal challenges in the near future are the inclusion of additional physics into the continuum modelling framework. In particular, we emphasise the need to incorporate phase-change behaviour into lithium transport models for the electrode particles (which is a key step in capturing hysteresis), and the necessity of building degradation mechanisms into the continuum modelling framework. We note that the latter is a very broad field, which includes several degradation mechanisms, and each of which has its own modelling challenges. These demands arise not only from requirements to formulate new models, but also in providing the methods to implement and parameterise these models so that they can be adopted by industry. Moreover, as computational capabilities improve over the coming years, we envisage that the use of more complex models will become more widespread. This applies at multiple levels: high-performance computing will enable the use of microscale models in battery design, and battery management systems with increased processing power should allow physics-based models to become standard components of battery management systems within industry.

\section*{Acknowledgements}
This work is supported by The Faraday Institution (EP/S003053/1 grant numbers FIRG003, FIRG015 and FIRG025). J.F. was supported by the Engineering and Physical Research Council (grant number EP/T000775/1). B.W. was supported by the Faraday Institution industrial fellowship (EP/S003053/1 grant number FIIF013).

The authors would like to thank Mr Amir Kosha Amiri for his graphical design expertise in constructing Figures \ref{fig:battery_sketch} to \ref{fig:sketch_SPMe}. These figures are available in the GitHub repository \url{https://github.com/FaradayInstitution/diagrams-battery-modelling}. The authors would also like to thank Prof David Howey for the useful feedback on the early draft.

\section*{CRediT author contributions statement}


\begin{itemize}
    
\item \textbf{Conceptualisation:} F.B.P., V.S., G.R


\item \textbf{Writing - Original Draft:} F.B.P., W.A., A.G., I.K., S.S., V.S., R.T., T.G.T, M.Z., J.S.E, J.M.F., G.R.

\item \textbf{Writing - Review \& Editing:} F.B.P., V.S., R.T., S.J.C., J.S.E., J.M.F., B.W., G.R.

\item \textbf{Visualisation:} F.B.P., A.M.B, T.G.T

\item \textbf{Supervision:} J.M.F, M.M., B.W., G.R.

\item \textbf{Project administration:} F.B.P.

\end{itemize}

\appendix

\section{List of symbols and acronyms}\label{sec:nomenclature}
Here we introduce a summary list of symbols and acronyms used throughout the paper, except the extensions in Section \ref{sec:extensions}. These require a notation very specific to each model and therefore hence it is kept self-contained within the section.

\noindent\textbf{Acroynyms}
\begin{description}[leftmargin=!, labelwidth=1.3cm, font=\normalfont]
    \item[DAE] Differential-algebraic equation
    \item[DFN] Doyle-Fuller-Newman
    \item[OCP] Open-circuit potential
    \item[ODE] Ordinary differential equation
    \item[PDE] Partial differential equation
    \item[SEI] Solid electrolyte interphase
    \item[SPM] Single Particle Model
    \item[SPMe] Single Particle Model with electrolyte dynamics
\end{description}

\noindent\textbf{Variables (with units)}
\begin{description}[leftmargin=!, labelwidth=1.3cm, font=\normalfont]
    \item[$c_k$] lithium concentration in the electrode particles \hfill mol m$^{-3}$ 
    \item[$c_\mre$] lithium ion concentration in the electrolyte \hfill mol m$^{-3}$ 
    \item[$\vb*{N}_\mre$] molar flux of ions in electrolyte \hfill mol m$^{-2}$ s$^{-1}$
    \item[$\vb*{i}_k$] current density in the electrodes \hfill A m$^{-2}$ 
    \item[$\vb*{i}_\mre$] current density in the electrolyte \hfill A m$^{-2}$ 
    \item[$j_k$] reaction current density \hfill A m$^{-2}$
    \item[$j_{k0}$] exchange current density \hfill A m$^{-2}$
    \item[$\phi_k$] electrode potential \hfill V
    \item[$\phi_\mre$]  electrolyte potential \hfill V
    \item[$T$] temperature \hfill K
    \item[$\eta_k$] overpotential at the electrode-electrolyte interface \hfill V
\end{description}

\noindent\textbf{Parameters and functions (with units)}
\begin{description}[leftmargin=!, labelwidth=1.3cm, font=\normalfont]
    \item[$b_k$] particle surface area per unit of volume \hfill m$^{-1}$
    \item[$c^{\max}_{k}$] maximum particle concentration \hfill mol m$^{-3}$
    \item[$c_{\mre 0}$] initial/rest lithium ion concentration in the electrolyte \hfill mol m$^{-3}$
    \item[$c_{k 0}$] electrode initial concentration \hfill mol m$^{-3}$
    \item[$c_p$] specific heat capacity \hfill J K$^{-1}$ kg$^{-1}$
    \item[$D_k$] lithium diffusivity in particle \hfill m$^2$ s$^{-1}$
    \item[$D_\mre$] lithium ion diffusivity in electrolyte \hfill m$^2$ s$^{-1}$
    \item[$h$] Heat transfer coefficient \hfill W m$^{-2}$ K$^{-1}$
    \item[$i_\mathrm{app}$] applied current density \hfill A m$^{-2}$
    \item[$K_k$] normalised reaction rate \hfill mol m$^{-2}$ s$^{-1}$
    \item[$L_k$] electrode and separator thicknesses \hfill m
    \item[$\dot Q_\mathrm{irr}$] irreversible heat source \hfill W m$^{-3}$
    \item[$\dot Q_\mathrm{rev}$] reversible heat source \hfill W m$^{-3}$
    \item[$R_k$] particle radius \hfill m
    \item[$t^+$] cation transference number \hfill -
    \item[$T_0$] initial temperature \hfill K
    \item[$T_\mathrm{amb}$] ambient temperature \hfill K
    \item[$U_k$] open-circuit potential (OCP) \hfill V
    \item[$\mathcal{B}_k$] transport efficiency / Inverse MacMullin number \hfill -
    \item[$\mathcal{K}$] thermal conductivity \hfill W m$^{-1}$ K$^{-1}$
    \item[$\varepsilon$] electrolyte volume fraction (porosity) \hfill -
    \item[$\theta$] volumetric heat capacity \hfill J K$^{-1}$ m$^{-3}$
    \item[$\rho$] density \hfill kg m$^{-3}$
    \item[$\sigma_\mre$] conductivity (electrolyte) \hfill S m$^{-1}$
    \item[$\sigma_k$] conductivity (electrode) \hfill S m$^{-1}$

\end{description}

\noindent\textbf{Constants (with units)}
\begin{description}[leftmargin=!, labelwidth=1.3cm, font=\normalfont]
    \item[$F$] Faraday constant \hfill C/mol
    \item[$R$] gas constant \hfill J/K/mol
\end{description}

\noindent\textbf{Subscripts}
\begin{description}[leftmargin=!, labelwidth=1.3cm, font=\normalfont]
    \item[e] in electrolyte
    \item[n] in negative electrode/particle (anode)
    \item[s] in separator
    \item[p] in positive electrode/particle (cathode)
    \item[$k$] in domain $k \in \{\mrn,\mrs,\mrp\}$
\end{description}

\noindent\textbf{Accents}
\begin{description}[leftmargin=!, labelwidth=1.3cm, font=\normalfont]
  \item[$\bar{}$] subdomain-averaged
  \item[$\tilde{}$] referred to the microscale
\end{description}

\bibliographystyle{abbrv}
\bibliography{references}

\end{document}